\documentclass[11pt,a4paper]{article}

\usepackage[english]{babel}

\usepackage[utf8]{inputenc}
\usepackage[T1]{fontenc}
\usepackage{lmodern}

\usepackage{amsmath,amssymb}
\usepackage{graphicx}
\usepackage{siunitx}
\usepackage{tikz}
\usetikzlibrary{arrows.meta,positioning,spy}
\usepackage{subcaption}
\usepackage{float}
\usepackage{pgfplots}
\pgfplotsset{compat=1.18}
\usepgfplotslibrary{groupplots}
\usepackage{paralist}
\usepackage{authblk}
\usepackage{hologo}
\usepackage[numbers,sort&compress]{natbib}

\usepackage{hyperref}
\hypersetup{
    pdfauthor={Federico Lo Presti},
    pdftitle={A Coupled CFD Framework for Combustor–Turbine Interaction in a Research Aeroengine}
}

\allowdisplaybreaks

\newcommand{\keywords}[1]{%
  \vspace{0.5em}
  \noindent\textbf{Keywords: }#1
}

\newenvironment{nomenclature}{%
  \section*{Nomenclature}
  \begin{description}
}{%
  \end{description}
}

\newcommand{\entry}[2]{\item[#1] #2}
\newcommand{\EntryHeading}[1]{\item[\textbf{#1}]}

\title{A Coupled CFD Framework for Combustor--Turbine Interaction\\
in a Research Aeroengine}

\author[1]{Federico Lo Presti\thanks{Corresponding author: federico.lopresti@rub.de}}
\author[2]{Pierre Vauquelin}
\author[1]{Jan Donndorf}
\author[1]{Francesca di Mare}
\author[2]{Xue-Song Bai}
\author[2]{Christer Fureby}

\affil[1]{Ruhr University Bochum, Bochum, Germany}
\affil[2]{Lund University, Lund, Sweden}

\date{} 

\begin{document}

\maketitle


\begin{abstract}
This work presents a fully coupled combustor--turbine simulation framework applied to the MYTHOS aeroengine, developed within the Horizon Europe project MYTHOS, aimed at assessing the impact of Sustainable Aviation Fuels (SAFs) and hydrogen on next-generation propulsion systems. 

The numerical setup features a dynamic, bidirectional coupling between a pressure-based solver with detailed finite-rate chemistry, deployed in the combustor, and a density-based turbomachinery solver employing tabulated thermochemistry for efficiency, used for the turbine. The coupling is realised through a flux-averaging methodology that ensures conservative exchange of flow quantities and allows flow in arbitrary directions across the interface. Previous validation steps of presented methodology have shown the viability of the approach and are also shortly reviewd. 
The paper focuses on the chemistry-handling strategy that guarantees thermochemical consistency between the two solvers. 

Coupled reacting simulations at cruise operating conditions demonstrate the capability of the framework to capture combustor-generated hot streaks transport and their influence on turbine aerothermal loading. 
Comparison with segregated simulations of the two components shows that coupling captures the highly unsteady temperature and flow distributions at the turbine inlet and across the blade rows. Whilst mean aerodynamic loading are essentially unchanged, a realistic circumferential variability in blade thermal loading can be observed in the coupled simulations, thus establishing a consistent foundation for future studies on the effects of alternative fuels on core engine components.
\end{abstract}

\begin{nomenclature}
\entry{$p$}{Static pressure [Pa]}
\entry{$\rho$}{Density [kg m $^{-3}$]}
\entry{$t$}{Time [s]}
\entry{$x_i$}{Spatial coordinate in direction $i$ [m]}
\entry{$u_i$}{Velocity component in direction $i$ [m s$^{-1}$]}
\entry{$e$}{Specific internal energy [J kg $^{-1}$]}
\entry{$h$}{Specific enthalpy [J kg $^{-1}$]}
\entry{$q_j$}{Heat flux component in direction $j$ [W m$^{-2}$]}
\entry{$Y_i$}{Mass fraction of species $i$ [--]}
\entry{$Z$}{Mixture fraction [--]}
\entry{$Z''^2$}{Mixture fraction variance [--]}
\entry{$Y_c$}{Progress variable [--]}
\entry{$C$}{Normalized progress variable, $C = Y_{\cdot}/\max(Y_{\cdot})$ [--]}
\entry{$D$}{Diffusivity [m$^2$ s$^{-1}$]}
\entry{$k$}{Turbulent kinetic energy [m$^2$ s$^{-2}$]}
\entry{$\omega$}{Specific dissipation rate [s$^{-1}$]}
\entry{$\mu_t$}{Turbulent dynamic viscosity [kg m$^{-1}$ s$^{-1}$]}
\entry{$Sc_t$}{Turbulent Schmidt number [--]}
\entry{$\tau_{ij}$}{Viscous stress tensor components [Pa]}
\entry{$F_i$}{Convective flux of quantity $i$}
\entry{$\bar{F}_{i,j}$}{Averaged flux of quantity $i$ across face $j$}

\EntryHeading{Subscripts and superscripts}
\entry{$(\cdot)''$}{Fluctuating quantity (Reynolds or Favredecomposition)}
\entry{$\overline{(\cdot)}$}{Reynolds-averaged quantity}
\entry{$\widetilde{(\cdot)}$}{Favre-averaged quantity}
\end{nomenclature}

\section{Introduction}

The need to decarbonize air transport is a major drive for the adoption of Sustainable Aviation Fuels (SAFs), kerosene-alternative liquid fuels derived from biomass, waste, or synthesis processes (e.g. Fischer-Tropsch) powered by renewable energy sources. 
While these offer a direct pathway to reduce life-cycle \(\mathrm{CO_2}\) emissions~\cite{Eurocontrol2021SAF}, non-\(\mathrm{CO_2}\) climate effects~\cite{Voigt2021,Lee2021} and improve local air quality~\cite{Moore2015}, their variable thermochemical properties significantly affect flame behavior, stability, and pollutant formation~\cite{Kang2019a,Peiffer2019,HolladayAbdullahHeyne2020,LinNurazaqWang2023,HarlassDischlKaufmann2024}.
Such differences affect the temperature and composition distribution at the combustor exit, with direct implications for the downstream High-Pressure Turbine (HPT) performance and durability. Accurate HPT flow prediction thus demands, more than ever, numerical tools capable of capturing the coupled interaction between combustion and turbine aerothermal response.  

Traditionally, combustors and turbines are analyzed separately to limit computational cost and, most importantly, because their underlying physics is better handled by different numerical methods. Combustion chambers operate at low Mach numbers, for which pressure-based solvers are more suitable. Capturing flame dynamics accurately requires scale-resolving turbulence modeling and detailed finite-rate chemistry, and the complex combustor geometry typically requires unstructured meshing.
In turbomachinery flows, by contrast, compressibility plays an important role. Density-based solvers on structured meshes, together with RANS or hybrid RANS/LES modeling for turbulence are often the methods of choice.

In most cases, the coupling of combustor and turbine is achieved by prescribing time-averaged or steady boundary conditions at the interfaces~\cite{papadogiannis2015,Tomasello2023}. 
While effective as a first approximation, this strategy neglects the transient and spatially non-uniform character of the flow at the combustor–turbine interface, where strong temperature and composition gradients directly affect turbine inlet conditions and, consequently, performance and durability.
These phenomena are collectively referred to as Combustor–Turbine Interaction (CTI)~\cite{McGuirk_2014}.

The flow exiting the combustor is characterized by temperature and species non-uniformities—commonly referred to as \emph{hot streaks}~\cite{butler1989redistribution}. Over the years, their dynamic has been extensively studied experimentally and numerically~\cite{butler1989redistribution,jenny2012hot,povey2005hot,rai1990navier,dorney1992hot,ong2008hot},
and the importance of reproducing their unsteady nature has been highlighted~\cite{dorney1992hot,takahashi1990unsteady}.
As hot streaks are convected through the vane and rotor passages, they interact with vane and blade rows and secondary flows, producing localized variations in heat transfer and sharp metal-temperature gradients. Their trajectory and persistence depend strongly on inlet swirl, turbulence intensity and the relative phase between combustor fluctuations and IGV leading edges. Many of these effects are neglected in single components simulations.

Several studies on combustor--turbine interaction presented simulations performed with a single solver including both components, but the turbine domain is typically restricted to the inlet guide vanes (IGV)~\cite{Cha2012,Klapdor2013,Jacobi2017,Govert2019,Tomasello2023}, as including the rotor demands resolving blade motion and unsteady blade–row interactions, which greatly increases the complexity of the numerical setup.

The most detailed studies rely on two distinct solvers, and various coupling strategies have been explored, either by interfacing different codes or by running multiple instances of the same solver with tailored numerical settings~\cite{legrenzi2017coupled}.
Early integrated studies~\cite{schlueter2004,schlueter2005} performed compressor/combustor, combustor/turbine, and full-engine simulations of a sector using URANS for turbomachinery and low-Mach LES for the combustor. A similar effort on a real engine sector~\cite{Turner2002,Turner2003,Turner2010} relied on exchanging radial boundary profiles, though mass-conservation issues were encountered. Other works explored direct solver-to-solver coupling, comparing steady single-component setups with coupled configurations~\cite{salvadori2012,papadogiannis2015,legrenzi2017coupled}. More recent studies~\cite{Miki2023}, where unsteady conditions were accounted for, showed marked differences between standalone and fully coupled simulations, including stronger temporal oscillations of turbine efficiency. Large-scale simulations have further emphasized the value of coupling: full-annulus LES of a fan–compressor–combustor system~\cite{PerezArroyo2021a,PerezArroyo2021b} and a coupled two-stage HPT with film and purge flows~\cite{Miki2025} demonstrated the improved accuracy obtained by integrated modelling of combustion, turbomachinery aerodynamics, and cooling.

This work advances previous research by introducing a coupling methodology which does not assume flow directionality across components, integrates finite-rate chemistry in the combustor with tabulated chemistry in the turbine, and is applied to a newly developed virtual test rig for SAF investigations. A detailed description of the coupling strategy and its verification is provided in~\cite{LoPresti2025}.

A central aspect of this study is the consistent treatment of chemistry across the combustor--turbine interface. Finite-rate chemistry is solved in the combustor to accurately capture the flame and heat-release dynamics, while the turbine simulation---where chemistry is assumed in the first instance to be frozen---relies on a precomputed table. The tabulation method, inspired by the Flamelet Generated Manifold (FGM) concept, parameterizes thermodynamic and transport quantities as a function of the mixture fraction~\(Z\), progress variable~\(Y_c\), mixture fraction variance~\(Z''^2\), and, to maintain thermodynamic consistency with the conservative density-based formulation, the specific internal energy~\(e\). This additional coordinate enables the temperature to be recovered consistently.

The introduction of an enthalpy- or energy-related coordinate within flamelet-based manifolds has been widely explored for representing heat-loss and other effects related to non-adiabatic conditions~\cite{fiorina2003modelling,ihme2008modeling,ketelheun2013heat,donini2013numerical,proch2015modeling}. Such extensions accommodate variations induced by wall cooling, dilution, or radiative losses that cannot be captured by adiabatic flamelets. 
In density-based solvers, and, generally, whenever an energy equation needs to be solved, discrepancies between the constant-pressure manifold and the thermodynamic state governed by the energy equation are often addressed through first-order corrections to the interpolated quantities~\cite{terrapon2009flamelet,vicquelin2011coupling,mittal2013flamelet,saghafian2015efficient,ma2017numerical}. These linear adjustments are effective when pressure fluctuations remain moderate, but become insufficient under large departures from the reference pressure. In those cases, additional coordinates---typically pressure or an equivalent thermodynamic variable---must be introduced into the chemical database to ensure consistency~\cite{vicquelin2011coupling}.

In our earlier work~\cite{lopresti2022numerical}, we analysed hot-streak propagation in a machine for which the combustor geometry was not accessible, using artificial unsteady temperature non-uniformities, imposed at turbine inlet. The present work overcomes these limitations by employing a fully resolved combustor–turbine configuration, representing the corresponding section of the MYTHOS Virtual Test Rig introduced in Sec.~\ref{sec:mythos-virtual-rig}, simulated at a representative operating point. The coupled approach relies on a solid validation of the combustion modeling techniques~\cite{sabelnikov2013lesmultiphase,vauquelin2025TARS}, the turbomachinery solver~\cite{lopresti2022numerical,karaefe2021,ziaja2021numerical,post2021a} and the coupling approach~\cite{LoPresti2025}.

\section{The MYTHOS Virtual Test Rig}
\label{sec:mythos-virtual-rig}
\begin{figure}[h]
    \centering
    \setlength{\fboxsep}{0pt}
    \fbox{%
        \includegraphics[width=\linewidth]{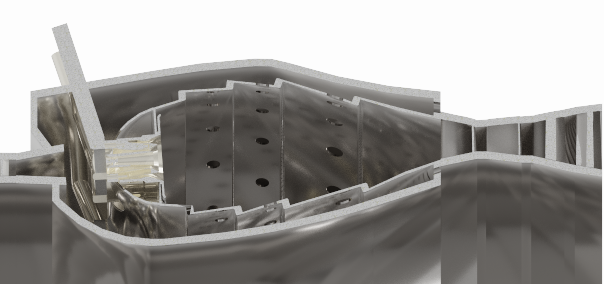}%
    }
    \caption{CAD illustration of combustor and turbine}
    \label{fig:geometry}
\end{figure}

The MYTHOS~\cite{MYTHOS2025} Virtual Test Rig (see~\cite{Donndorf2025,donndorf2025baseline,vauquelin2025optimal,lopresti2025virtual, vauquelin2026design}) serves as a dedicated research platform for investigating hybrid, low-emission propulsion systems compatible with Sustainable Aviation Fuels (SAFs) and hydrogen. Conceived for a medium-range aircraft application, it represents the core of a $\approx 100$ kN class turbofan engine and integrates a High-Pressure Compressor (HPC), an annular combustor, and a High-Pressure Turbine (HPT). Each component has been specifically designed to capture the influence of alternative fuels on aerothermal performance, combustion characteristics, emissions, and inter-component interactions under realistic operating conditions.
Figure~\ref{fig:geometry} shows the geometry of the two components under investigation in this study.

The overall engine architecture was established through a zero-dimensional thermodynamic cycle 
model~\cite{Donndorf2025}, which defined key design targets such as pressure ratios, mass flows, and the thermodynamic state of the working fluid at each engine station. These parameters informed the detailed CFD-based design of the combustor and the meanline and three-dimensional aerodynamic design of the turbomachinery components. The resulting virtual platform enables high-fidelity numerical investigations of the coupled core components, providing physical insight into SAF and hydrogen operations. The data generated will support the development of Reduced Order Models (ROMs) for each component, to be integrated into a holistic digital framework of the engine, enabling rapid and fuel-specific assessments of performance and pollutant formation.

\section{Mathematical model}
\label{sec:mathematical-model}
\subsection{Combustor}
Combustor simulations rely on the pressure-based, compressible reacting-flow framework of \textit{OpenFOAM}~\cite{weller1998tensorial}. The governing equations follow the unsteady Favre-averaged Navier--Stokes formulation for compressible reacting flows~\cite{poinsot2005turbulentcombustion,menon2010computationalcombustion}:
\begin{equation}
\partial_t \bar{\rho} + \nabla\!\cdot(\bar{\rho}\tilde{\mathbf{u}}) = \dot{\rho}_s,
\label{eq:RANS_rho}
\end{equation}
\begin{equation}
\partial_t(\bar{\rho}\tilde{\mathbf{u}}) + \nabla\!\cdot(\bar{\rho}\tilde{\mathbf{u}}\tilde{\mathbf{u}})
= -\nabla \bar{p}
+ \nabla\!\cdot\left(\bar{\boldsymbol{\tau}}
      - \bar{\rho}\widetilde{\mathbf{u}''\mathbf{u}''}\right)
+ \dot{\mathbf{m}}_s,
\label{eq:RANS_mom}
\end{equation}
\begin{equation}
\partial_t(\bar{\rho}\tilde{H}) + \nabla\!\cdot(\bar{\rho}\tilde{\mathbf{u}}\tilde{H})
= \partial_t\bar{p}
+ \nabla\!\cdot\!\left(\bar{\boldsymbol{\tau}}\tilde{\mathbf{u}}
      + \mathbf{q}
      - \bar{\rho}\widetilde{\mathbf{u}''h''}\right)
+ \sum_i \dot{\omega}_i h_{f,i}
+ \dot{q}_s,
\label{eq:RANS_energy}
\end{equation}
\begin{equation}
\partial_t(\bar{\rho}\tilde{Y}_i) + \nabla\!\cdot(\bar{\rho}\tilde{\mathbf{u}}\tilde{Y}_i)
= \nabla\!\cdot\left(\bar{\rho}D_i\nabla\tilde{Y}_i
      - \bar{\rho}\widetilde{\mathbf{u}''Y_i''}\right)
+ \dot{\omega}_i + \dot{\rho}_{s,i}.
\label{eq:RANS_species}
\end{equation}

Here, $\bar{\rho}$ is the mean density, $\tilde{\mathbf{u}}$ the Favre velocity, $\bar{p}$ the mean pressure, and $\tilde{Y}_i$ the Favre mass fraction of species $i$. The total enthalpy is $\tilde{H} = \tilde{h}_s + |\tilde{\mathbf{u}}|^2/2$, with $\tilde{h}_s$ the sensible enthalpy. The viscous stress tensor is $\bar{\boldsymbol{\tau}}$ and the Reynolds stresses $\widetilde{\mathbf{u}''\mathbf{u}''}$ are closed with the Boussinesq approximation using the $k$--$\omega$ SST model. Molecular diffusion uses $D_i$ and $\mathbf{q}=-\kappa\nabla\tilde{T}$ is the conductive heat flux. The terms $\dot{\rho}_s$, $\dot{\rho}_{s,i}$, $\dot{\mathbf{m}}_s$ and $\dot{q}_s$ represent mass, species, momentum and energy exchange between the gas and the spray phase. Chemical production rates are $\dot{\omega}_i$ and $h_{f,i}$ are formation enthalpies. An ideal-gas equation of state is used, together with Fourier heat conduction and Fickian diffusion~\cite{poinsot2005turbulentcombustion}. The $k$--$\omega$ SST turbulence model provides the eddy viscosity
\begin{equation}
\mu_t = \bar{\rho}\,a_1\,\frac{k}{\max(a_1\omega, S F_2)},
\end{equation}
with $k$ the turbulent kinetic energy, $\omega$ the specific dissipation rate, $S$ the strain-rate magnitude, and $F_2$ the SST blending function. Turbulent scalar transport uses a gradient--diffusion hypothesis, with $D_{\mathrm{eff}} = D + \mu_t/(\bar{\rho}Sc_t)$ and $\kappa_{\mathrm{eff}} = \kappa + \mu_t/Pr_t$, using $Sc_t=1.0$ and $Pr_t=0.85$.

Combustion is treated with Finite-Rate Chemistry (FRC) based on Arrhenius kinetics~\cite{poinsot2005turbulentcombustion}. A PaSR-type closure~\cite{sabelnikov2013lesmultiphase} accounts for turbulence--chemistry interaction by weighting the homogeneous reaction rate with mixing and chemical time scales, consistently reflecting the local Damk{\"o}hler number~\cite{damkohler1936einfluesse}. The Jet~A spray is modeled using a four-way-coupled Discrete Droplet Method (DDM), where Lagrangian parcels follow the particle--fluid formulation of Dukowicz~\cite{dukowicz1980particlemodel}, with stochastic secondary breakup and parcel management following~\cite{apte2003atomizingspray,reitz1987atomizationmechanisms,reitz1986dropbreakup} and evaporation based on Ranz--Marshall correlations~\cite{ranz1952evaporationdrops}. Gas-phase chemical kinetics are modeled using the reduced Z79 mechanism~\cite{zettervall2021thesis,AkerblomZettervallFureby} combined with the Yoshikawa--Reitz NO$_x$ mechanism~\cite{yoshikawa2009noxmechanism}, yielding a 37-species, 98-reaction system. The mechanism reproduces laminar flame speeds and ignition delay times of Jet~A~\cite{Hui2013,Kumar2011} and agrees with detailed and skeletal mechanisms (CRECK~\cite{CRECK} and HyChem~\cite{HyChemJetA}).

\subsection{Turbine}
Turbine simulations are performed with the in-house density-based solver \textit{SharC}~\cite{post2021a, ziaja2021numerical, karaefe2021, lopresti2022numerical}, a finite-volume, multi-block structured solver for compressible multi-species, multi-phase flows.
The single-phase, Favre-averaged Navier--Stokes equations (Eqs. ~\eqref{eq:continuity}--~\eqref{eq:energy})
are complemented by the transport equations for the mixture fraction $Z$ (Eq. ~\eqref{eq:mixturefraction}), a progress variable $Y_c$ (Eq. ~\eqref{eq:progressvariable}) and the mixture fraction variance $Z''$.
More detail on the tabulation approach for the representation of the mixture composition and calculation of thermodynamic and transport properties will be introduced in Sec.~\ref{subsubsec:tabulated-chemistry}.

\begin{equation}
\frac{\partial \overline{\rho}}{\partial t} + \frac{\partial (\overline{\rho} \widetilde{u}_i)}{\partial x_i} = 0
\label{eq:continuity}
\end{equation}

\begin{equation}
\frac{\partial (\overline{\rho} \widetilde{u}_j)}{\partial t} + \frac{\partial (\overline{\rho} \widetilde{u}_i \widetilde{u}_j)}{\partial x_i} = -\frac{\partial \overline{p}}{\partial x_j} + \frac{\partial}{\partial x_i} \left( \overline{ \tau_{ij}} - \overline{\rho} \widetilde{u}_i'' \widetilde{u}_j'' \right)
\label{eq:momentum}
\end{equation}

\begin{equation}
\begin{aligned}
\frac{\partial}{\partial t} \left[ \overline{\rho} \left( \widetilde{e} + \frac{\widetilde{u}_i \widetilde{u}_i}{2} + \frac{\widetilde{u_i'' u_i''}}{2} \right) \right] 
+ \frac{\partial}{\partial x_j} \left[ \overline{\rho} \widetilde{u}_j \left( \widetilde{h} + \frac{\widetilde{u}_i \widetilde{u}_i}{2} + \frac{\widetilde{u_i'' u_i''}}{2} \right) \right] \\
= \frac{\partial}{\partial x_j} \left[ -\overline{q}_j - \overline{\rho} \widetilde{u_j'' h''} \right] 
+ \frac{\partial}{\partial x_j} \left[ \widetilde{u}_i \left( \overline{\tau_{ij}} - \overline{\rho} \widetilde{u_i'' u_j''} \right) \right]
\end{aligned}
\label{eq:energy}
\end{equation}

\begin{equation}
\frac{\partial (\overline{\rho} \,\widetilde{Z})}{\partial t}
+ \frac{\partial (\overline{\rho} \,\widetilde{u}_i \,\widetilde{Z})}{\partial x_i}
=
\frac{\partial}{\partial x_i}
\left(
\overline{\rho D_Z \frac{\partial Z}{\partial x_i}}
- \overline{\rho} \,\widetilde{u_i'' Z''}
\right)
\label{eq:mixturefraction}
\end{equation}

\begin{equation}
\frac{\partial (\overline{\rho} \,\widetilde{Y}_C)}{\partial t}
+ \frac{\partial (\overline{\rho} \,\widetilde{u}_i \,\widetilde{Y}_C)}{\partial x_i}
=
\frac{\partial}{\partial x_i}
\left(
\overline{\rho D_C \frac{\partial Y_C}{\partial x_i}}
- \overline{\rho} \,\widetilde{u_i'' Y_C''}
\right)
\label{eq:progressvariable}
\end{equation}

\begin{equation}
\begin{aligned}
\frac{\partial \left( \overline{\rho}\,\widetilde{Z''^2} \right)}{\partial t}
+ \frac{\partial}{\partial x_i}
\left( \overline{\rho}\,\widetilde{u}_i\,\widetilde{Z''^2} \right)
&= 
\frac{\partial}{\partial x_i}
\left(
\overline{\rho\,D_Z\,\frac{\partial Z''^2}{\partial x_i}}
- 
\overline{\rho}\,\widetilde{u_i'' Z''^2}
\right) \\
&\quad 
- 2\,\overline{\rho}\,\widetilde{u_i'' Z''}\,
\frac{\partial \widetilde{Z}}{\partial x_i} 
- 2\,
\overline{\rho\,D_Z\,
\frac{\partial Z''}{\partial x_i}
\frac{\partial Z''}{\partial x_i}}
\end{aligned}
\label{eq:mixturefractionvariance}
\end{equation}

In~\eqref{eq:energy}, molecular diffusion and the turbulent transport of turbulent kinetic energy are neglected, while the energy flux is modeled using Fourier’s law and Fick’s law, assuming equal diffusivity for all species and a unity Lewis number.

The unclosed terms are modeled for the present application using an (Unsteady)-RANS approach with the \(k\text{-}\omega\) SST turbulence model as well.

Turbulent scalar fluxes 
$\widetilde{u_i''\phi''}$ are closed through the gradient–diffusion 
hypothesis so that an effective diffusivity 
$D_{\text{eff}} = D + \mu_t/(\bar{\rho}Sc_t)$ is used. 
In Eq.~\eqref{eq:mixturefractionvariance}, 
the production term becomes 
$2(\mu_t/Sc_t)\,\partial_i\tilde{Z}\,\partial_i\tilde{Z}$, and the 
dissipation is modelled as 
$C_x\,\bar{\rho}(\varepsilon/k)\,\widetilde{Z''^2}$~\cite{terrapon2009flamelet}.

\subsubsection{Tabulated chemistry}
\label{subsubsec:tabulated-chemistry}
\begin{figure}[h]
    \centering
    \includegraphics[width=\linewidth]{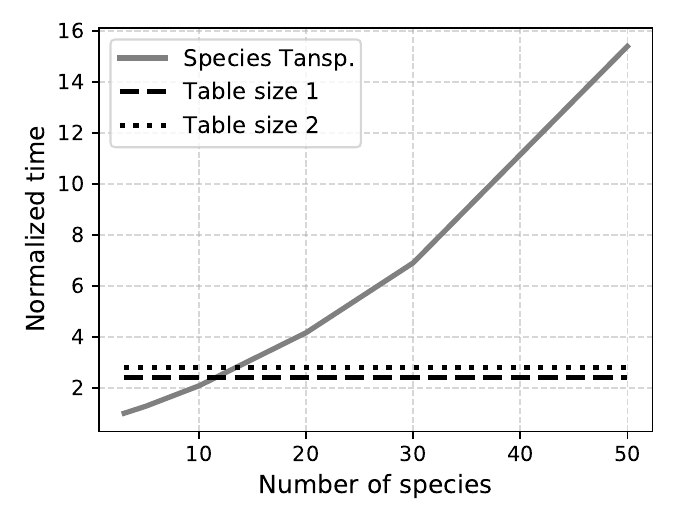}
    \caption{Computational cost comparison: species transport vs tabulation}
    \label{fig:mixture-cost}
\end{figure}

As already introduced, little to no further chemical reactions occur in the turbine domain, yet the gas composition and thermodynamic state remain influenced by the upstream combustion process.
To represent these effects efficiently and maintain thermodynamic consistency of the coupled simulations, a \emph{tabulated chemistry approach} based on precomputed thermochemical data is adopted, avoiding the solution of a transport equation for each of the species included in the chosen reaction mechanism, the cost of which would not be justified by an increased accuracy. 
Figure~\ref{fig:mixture-cost} compares the computational cost of simulations with explicit species transport, evaluated for an increasing number of species, against the tabulated approach at two table resolutions ($\approx 14 \text{k}$ and $\approx 42\text{k}$ points). As expected, the cost of species transport grows with the number of species considered (size of the combustor reaction mechanism), whereas the tabulated method shows only a modest increase with table resolution. Because table size is linked to the desired interpolation accuracy rather than to the number of species represented, the approach remains efficient even when very large mechanisms are used to describe a more complex chemistry in the combustion chamber. In the present setup, the tabulated approach becomes advantageous when more than $\approx 12$ species are considered.

The chemistry table is generated from a series of freely propagating premixed laminar flames computed with \textsc{Cantera}~\cite{Cantera2025}. The solution fields are stored as functions of the mixture fraction~\(Z\), the normalized progress variable, defined as \(C = Y_{\mathrm{CO_2}}/\max\!\left(Y_{\mathrm{CO_2}}(Z,e)\right)\) and the specific internal energy~\(e\). 
As typical~\cite{terrapon2009flamelet,saghafian2015efficient},
presumed PDFs are employed to represent turbulence--chemistry interaction: a $\beta$ distribution for the mixture fraction $Z$ and $\delta$ distributions for the progress variable $C$. A $\delta$ distribution is assumed here for the internal energy as well.

During the simulation, thermochemical quantities are obtained from averaged quantities by interpolation within the four-dimensional manifold:  
\begin{equation}
    \phi = \phi \left(\widetilde{Z}, \widetilde{Z''^2}, \widetilde{C}, \widetilde{e} \right)
\end{equation}

\begin{figure}
    \centering
    \includegraphics[width=\linewidth]{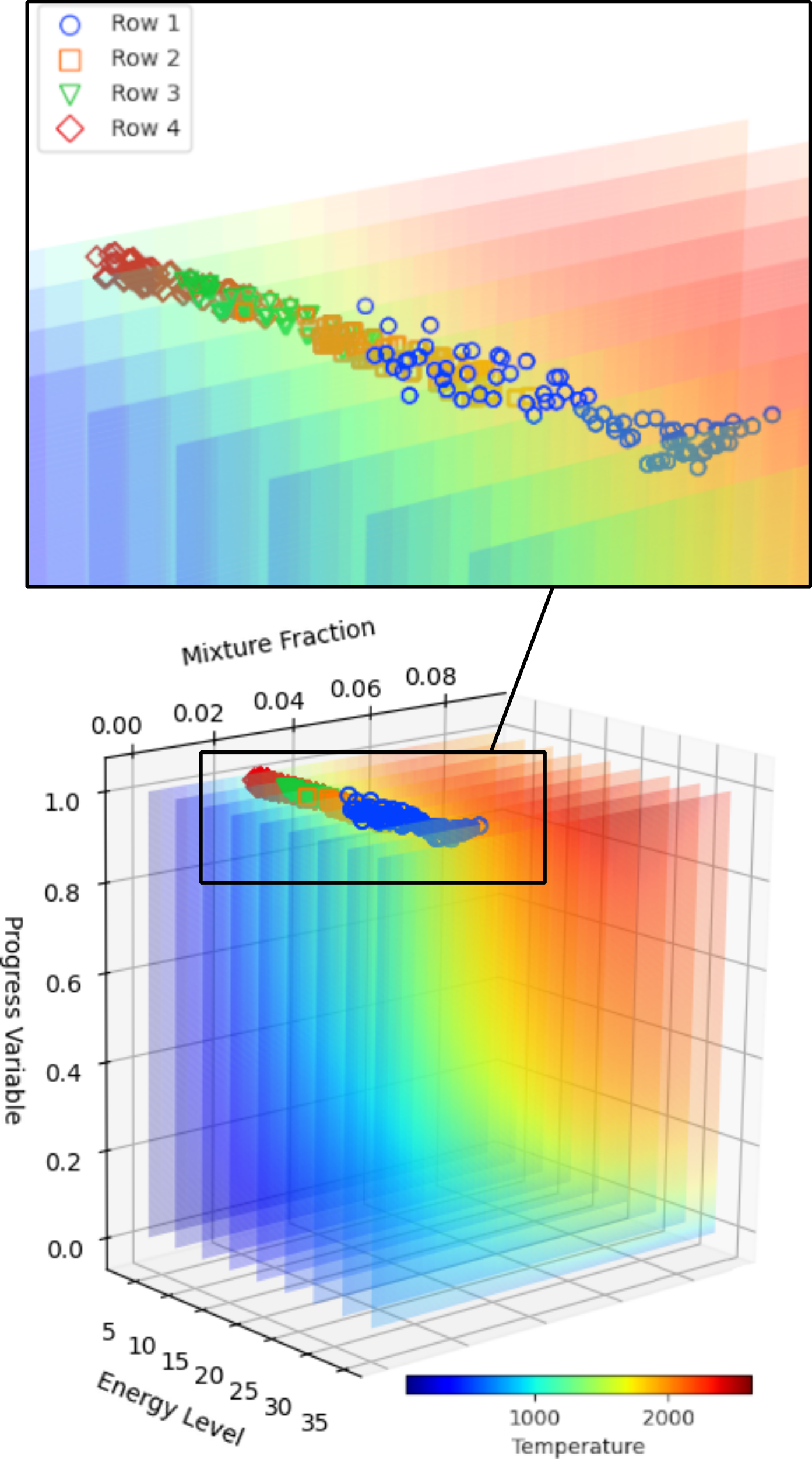}
    \caption{Look up table (Temperature) and tabulation-coordinate samples throughout the turbine domain}
    \label{fig:table-position}
\end{figure}

The table used in the following numerical tests and in the application of Sec.~\ref{subsec:results-reacting}, shown in Fig.~\ref{fig:table-position}, consists of 14 levels in \(Z\), 201 levels in \(C\), 39 levels in \(e\), and is integrated using 5 presumed--PDF levels for \(Z''^{2}\).

\begin{figure}[h]
    \centering
    \includegraphics[width=\linewidth]{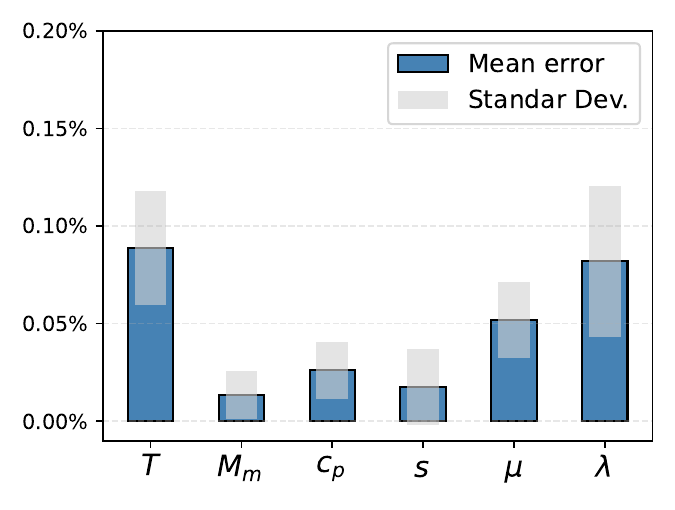}
    \caption{Mean and std. deviation of mapping error at interface}
    \label{fig:mapping-error}
\end{figure}

\begin{figure}[h]
    \centering
    \includegraphics[width=\linewidth]{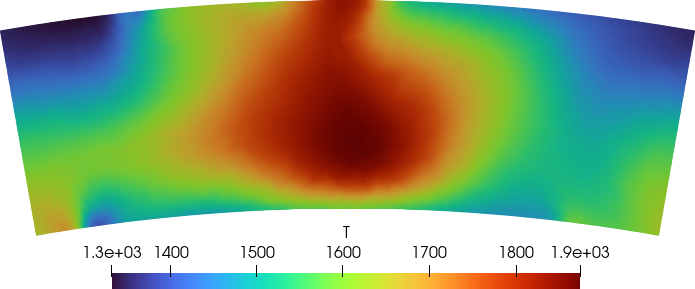}
    \caption{Temperature distribution at combustor outlet}
    \label{fig:combustor-exit}
\end{figure}

At the combustor--turbine interface, the tabulation coordinates \((Z,\, Z''^{2},\, Y_{c},\, e)\) are reconstructed directly from the local mixture composition \(Y_i\) \((i = 1,\ldots,N)\). Figure~\ref{fig:mapping-error} reports the mean and standard deviation of the difference between the quantities of interest for the turbine simulation, evaluated from the table and computed directly from the local mixture composition and temperature field at combustor outlet, a snapshot of which from the current simulation is shown in Fig.~\ref{fig:combustor-exit}. The mapping error remains on the order of \(0.1\%\), indicating an excellent consistency between the reconstructed and fully evaluated thermochemical states.

As a preliminary verification step towards the fully coupled simulations, a steady RANS computation of the isolated turbine was carried out using the combustor outlet field of Fig.~\ref{fig:combustor-exit} as a steady, two-dimensional inlet boundary condition. Figure~\ref{fig:table-position} presents a scatter plot of the tabulation coordinates sampled throughout the turbine, separated by blade row. As the expansion proceeds, the mixture shifts towards lower energy levels,
confirming the need to account for more energy levels in the tabulation strategy.
Also, the figures shows how the variability in mixture fraction and progress variable is greatest in the first row as a consequence of the inhomogeneous mixture, and tends, as expected, towards a single, mixed state at turbine exit.
The mixture entering the turbine lies on the lean side of the flamelet manifold (\(Z < Z_{\mathrm{st}}\)) and corresponds to nearly complete combustion (\(C \approx 1\)). As a consequence, the thermochemical states populate only a small region of the tabulated domain, indicating that the table required for the turbine simulation could, in principle, be significantly reduced in size.

\subsection{Solver Coupling}
The two solvers introduced in Sec.~\ref{sec:mathematical-model} are coupled in real time through an MPI-based data-exchange layer.

As discussed in~\cite{LoPresti2025}, the flux-based formulation ensures conservativeness at the interface compared to prescribing primitive variables directly, and the averaging procedure makes the boundary behave as a proper numerical interface, enabling flow in arbitrary directions being exchanged.

In the current configuration, both solvers employ the same turbulence modelling approach and no special treatment is required at the interface to exchange turbulent fluctuations.

\section{Results and discussion}
\subsection{Numerical setup}
The whole engine configuration has been briefly introduced 
in Sec.~\ref{sec:mythos-virtual-rig}, while the solvers, turbulence and combustion modeling are described in Sec.~\ref{sec:mathematical-model}. 
Computations of combustor and turbine are performed on a $20^{\circ}$ sector, corresponding to one burner of the annular combustion chamber, with cyclic periodic boundary conditions applied on the lateral faces to preserve azimuthal flow structures while limiting computational cost.

The combustor geometry includes the down-scaled TARS injector \cite{TARSLiGutmark2005,vauquelin2025TARS} and the full flame tube with mixing, dilution and cooling holes. At the combustor inlet, a uniform total temperature and mass-flow rate consistent with the engine Cruise conditions~\cite{Donndorf2025} are imposed. All solid walls are treated as no-slip with wall functions used for the near-wall turbulence treatment in the $k$--$\omega$ SST model. The heat shield is fixed at 1300\,K, following modern thermal-barrier-coating temperature limits. On the flame-facing surfaces of the liner, stage-specific one-dimensional heat-transfer coefficients are applied. These coefficients are derived from Nusselt numbers estimations from the air mass-flow distribution and Reynolds numbers obtained from non-reactive simulations.

At turbine outlet, a static-pressure condition is applied, with the value corresponding to the discharge pressure of the HPT, while solid surfaces are modelled as adiabatic walls and a Spalding wall function is used.

\begin{figure}[h]
\centering
\includegraphics[width=\linewidth]{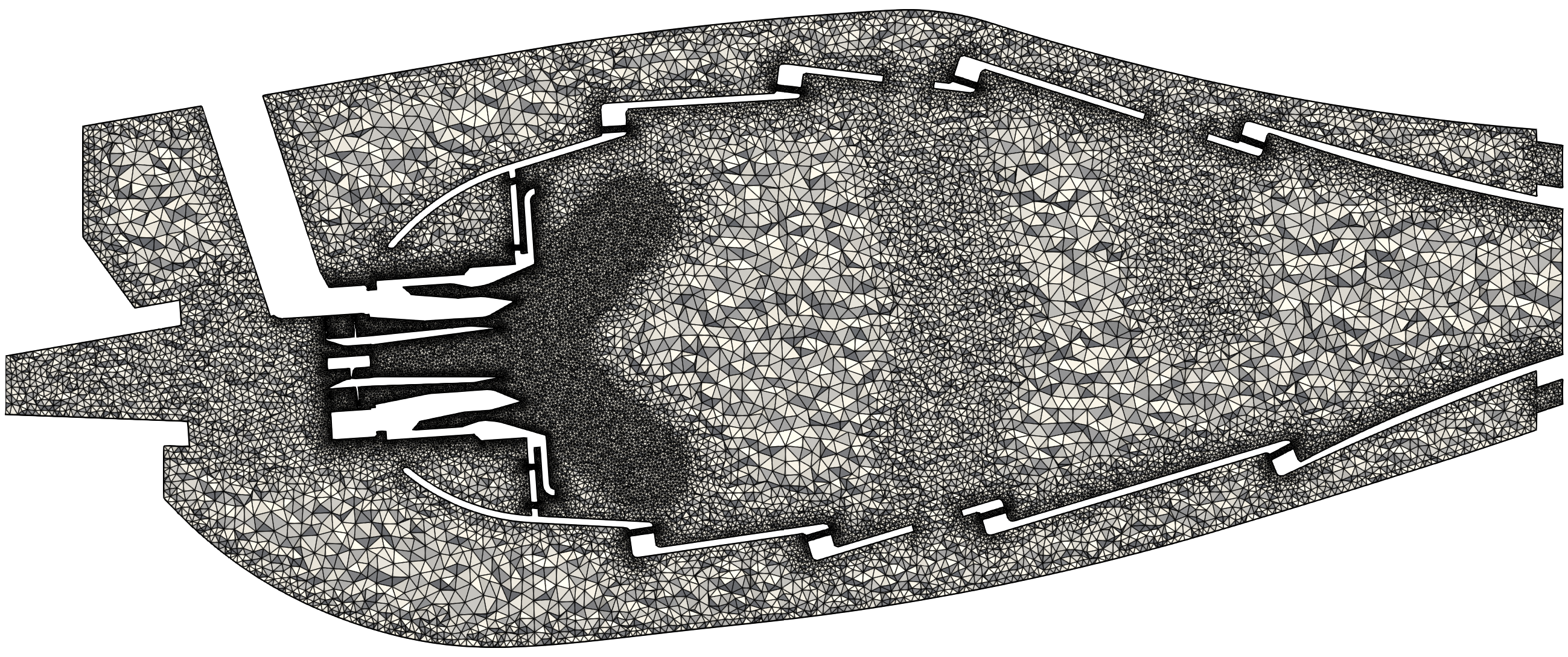}
\caption{Center-plane slice across the combustor grid.}
\label{fig:MeshCombustor}
\end{figure}

\begin{figure}[h]
    \centering
    \setlength{\fboxsep}{0pt}
    \fbox{%
        \includegraphics[width=\linewidth]{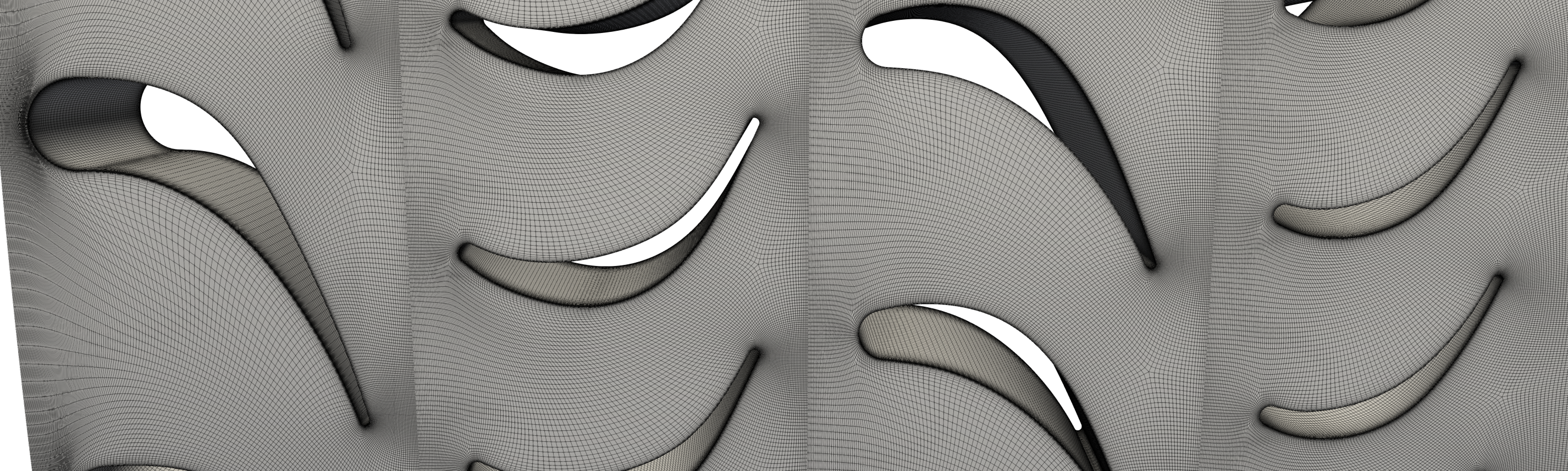}%
    }
    \caption{Detail of turbine mesh}
    \label{fig:turbine-mesh}
\end{figure}

The computational domain of the combustor is discretized using an unstructured mesh composed of approximately 12 million tetrahedral elements, refined in the injector region, the recirculation zone and the vicinity of the flame front, as shown in Fig.~\ref{fig:MeshCombustor}.
The multi-block structured mesh for the turbine, whose detail over the airfoils is shown in Fig.~\ref{fig:turbine-mesh} comprises $\approx 26.7\times10^6$ cells, with refinement around solid surfaces and leading and trailing edges in particular.


The combustor time step is set to $6.25\times10^{-8} \mathrm{s}$, to keep the CFL number safely below one (0.6 on average).
In the turbine, dual time-stepping is employed with a physical time step of $2.5\times10^{-7} \mathrm{s}$.
Flux exchange across the coupled interface is therefore performed at each physical turbine time step and every four combustor steps.

Both solvers are initialized from fully developed standalone solutions, in which uniform boundary conditions replace the coupling interface. The coupled simulation is then advanced for a total of 15 ms, corresponding to approximately four turbine revolutions. An initial transient of 4.6 ms is discarded to allow the development of the coupled flow field, after which statistics are accumulated over roughly 10 ms. This averaging period provides decent convergence of first-order statistics, allowing a consistent evaluation of averaged quantities.

Simulations are carried out on 8 CPU nodes (1024 cores) of the LUMI supercomputer, enabled by the EuroHPC Regular Access Project EHPC-REG-2024R02-188~\cite{EuroHPC}.
The computational cost is approximately $1.5\times10^{5}$ CPU-hours for a simulated time corresponding to one turbine revolution.
The computational load is balanced across nodes to ensure that both solvers reach the coupling point simultaneously, minimizing idle time during data exchange.




\subsection{Reacting flow}
\begin{figure}[h]
    \centering
    \def\svgwidth{\linewidth}
\begingroup%
  \makeatletter%
  \providecommand\color[2][]{%
    \errmessage{(Inkscape) Color is used for the text in Inkscape, but the package 'color.sty' is not loaded}%
    \renewcommand\color[2][]{}%
  }%
  \providecommand\transparent[1]{%
    \errmessage{(Inkscape) Transparency is used (non-zero) for the text in Inkscape, but the package 'transparent.sty' is not loaded}%
    \renewcommand\transparent[1]{}%
  }%
  \providecommand\rotatebox[2]{#2}%
  \newcommand*\fsize{\dimexpr\f@size pt\relax}%
  \newcommand*\lineheight[1]{\fontsize{\fsize}{#1\fsize}\selectfont}%
  \ifx\svgwidth\undefined%
    \setlength{\unitlength}{750bp}%
    \ifx\svgscale\undefined%
      \relax%
    \else%
      \setlength{\unitlength}{\unitlength * \real{\svgscale}}%
    \fi%
  \else%
    \setlength{\unitlength}{\svgwidth}%
  \fi%
  \global\let\svgwidth\undefined%
  \global\let\svgscale\undefined%
  \makeatother%
  \begin{picture}(1,0.84006665)%
    \lineheight{1}%
    \setlength\tabcolsep{0pt}%
    \put(0,0){\includegraphics[width=\unitlength,page=1]{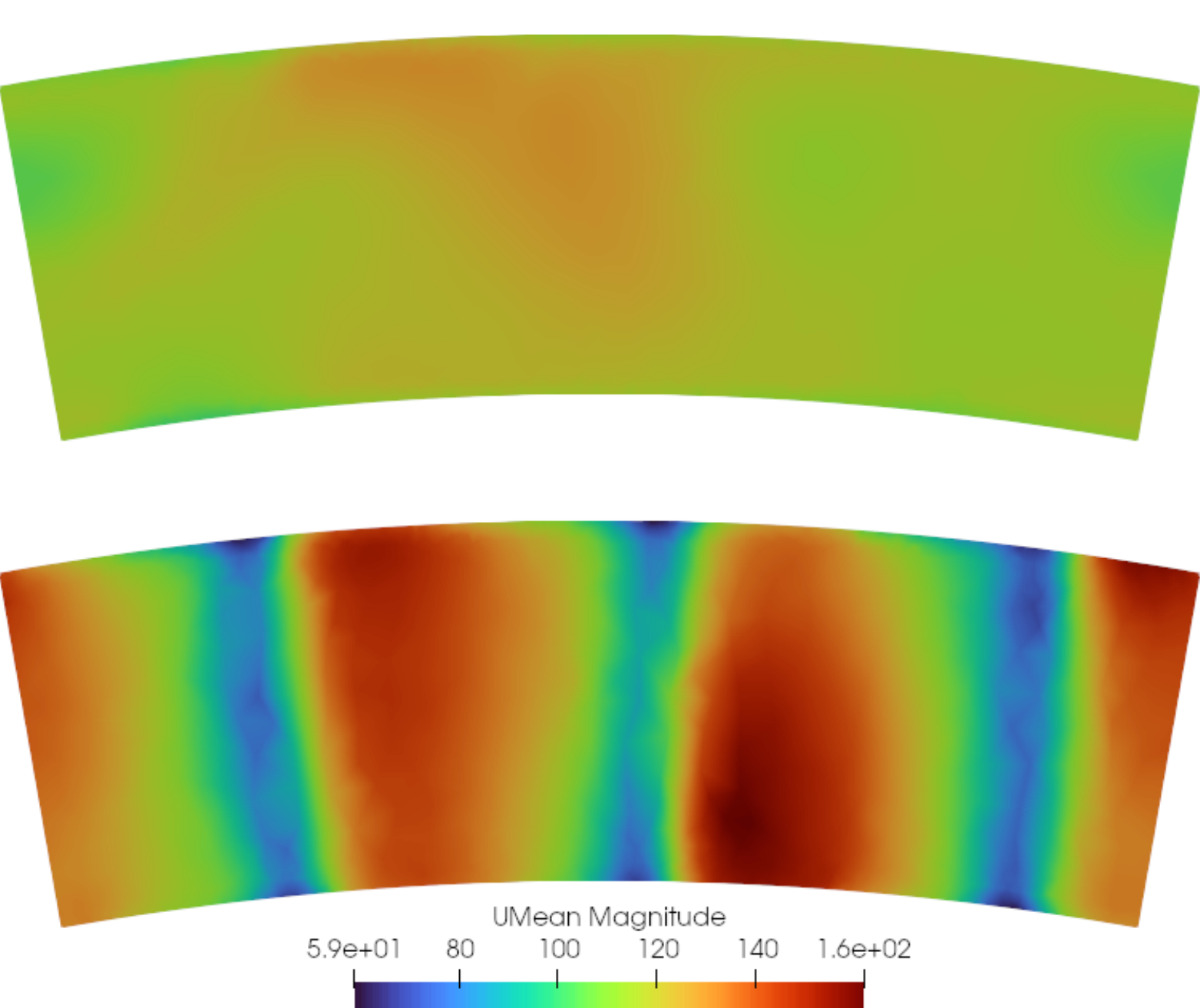}}%
    \put(0.76790473,0.40290457){\color[rgb]{0,0,0}\makebox(0,0)[lt]{\lineheight{1.25}\smash{\begin{tabular}[t]{l}COUPLED\end{tabular}}}}%
    \put(0.73118475,0.81106665){\color[rgb]{0,0,0}\makebox(0,0)[lt]{\lineheight{1.25}\smash{\begin{tabular}[t]{l}STANDALONE\end{tabular}}}}%
  \end{picture}%
\endgroup%

    \caption{Comparison of velocity magnitude field at combustor outlet: standalone vs coupled}
    \label{fig:umag-out-compare}
\end{figure}

The step towards a more realistic numerical setup, besides having significant effects on the turbine results, also affects the flow field in the combustion chamber. Although the turbine inlet vanes are choked, as it is common design practice, replacing the uniform-pressure outlet condition alters the pressure distribution at the combustor–turbine interface. Figure~\ref{fig:umag-out-compare} compares the velocity magnitude at the combustor outlet for the standalone and coupled simulations and shows how the modified pressure field leads to local deceleration near the IGV stagnation regions and acceleration within the passages.

\begin{figure*}[t]
    \centering
    \setlength{\fboxsep}{0pt}

    \begin{subfigure}[t]{0.84\linewidth}
        \centering
    \fbox{%
        \includegraphics[
            width=\linewidth,
            trim={0 420 352 530},      
            clip
        ]{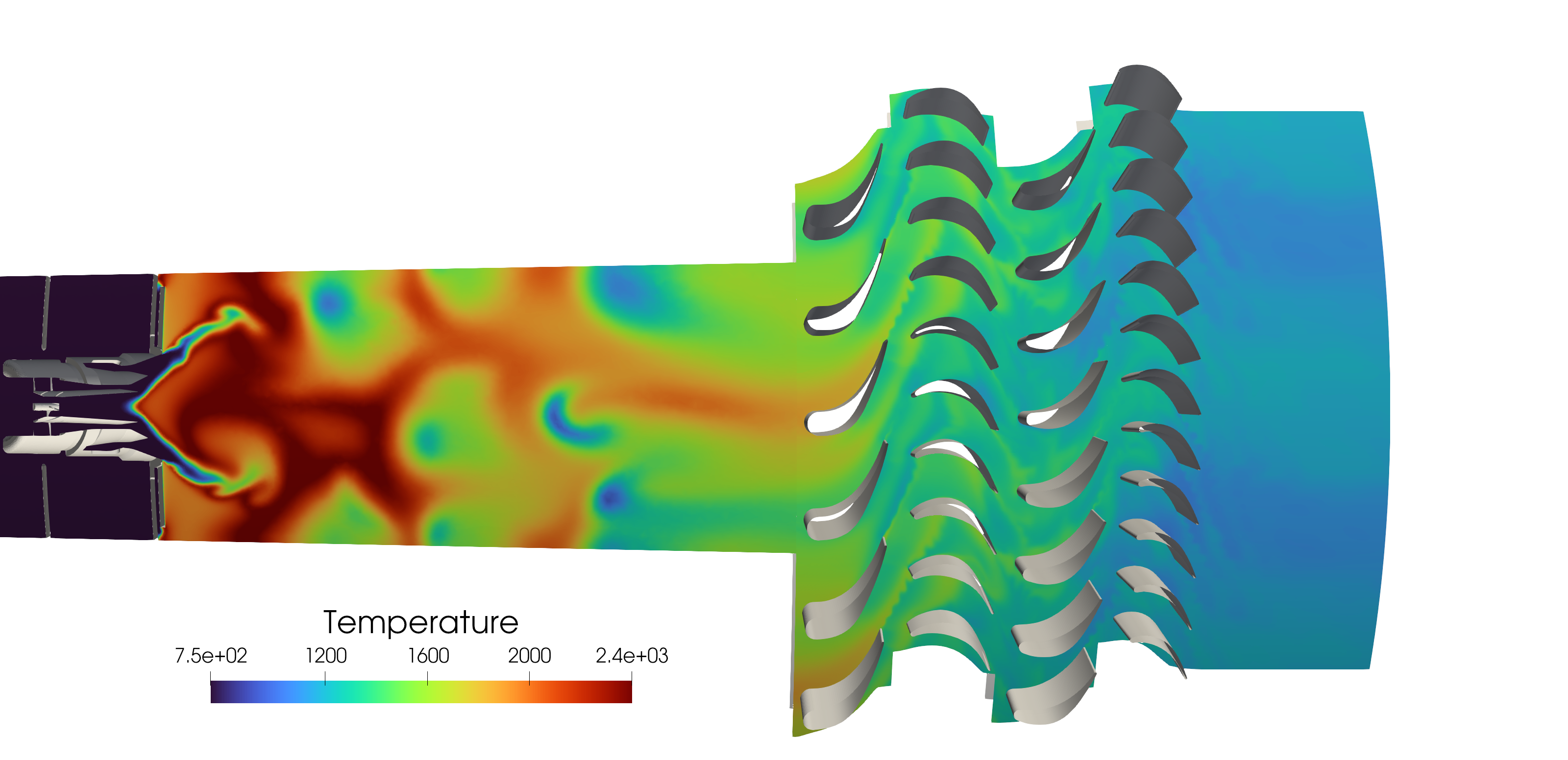}
        }
        \caption{$t/t_{rot} = 1.625$}
    \end{subfigure}
    \hfill
            \includegraphics[
            width=0.11\linewidth,
            trim={2868 440 20 550},
            clip
        ]{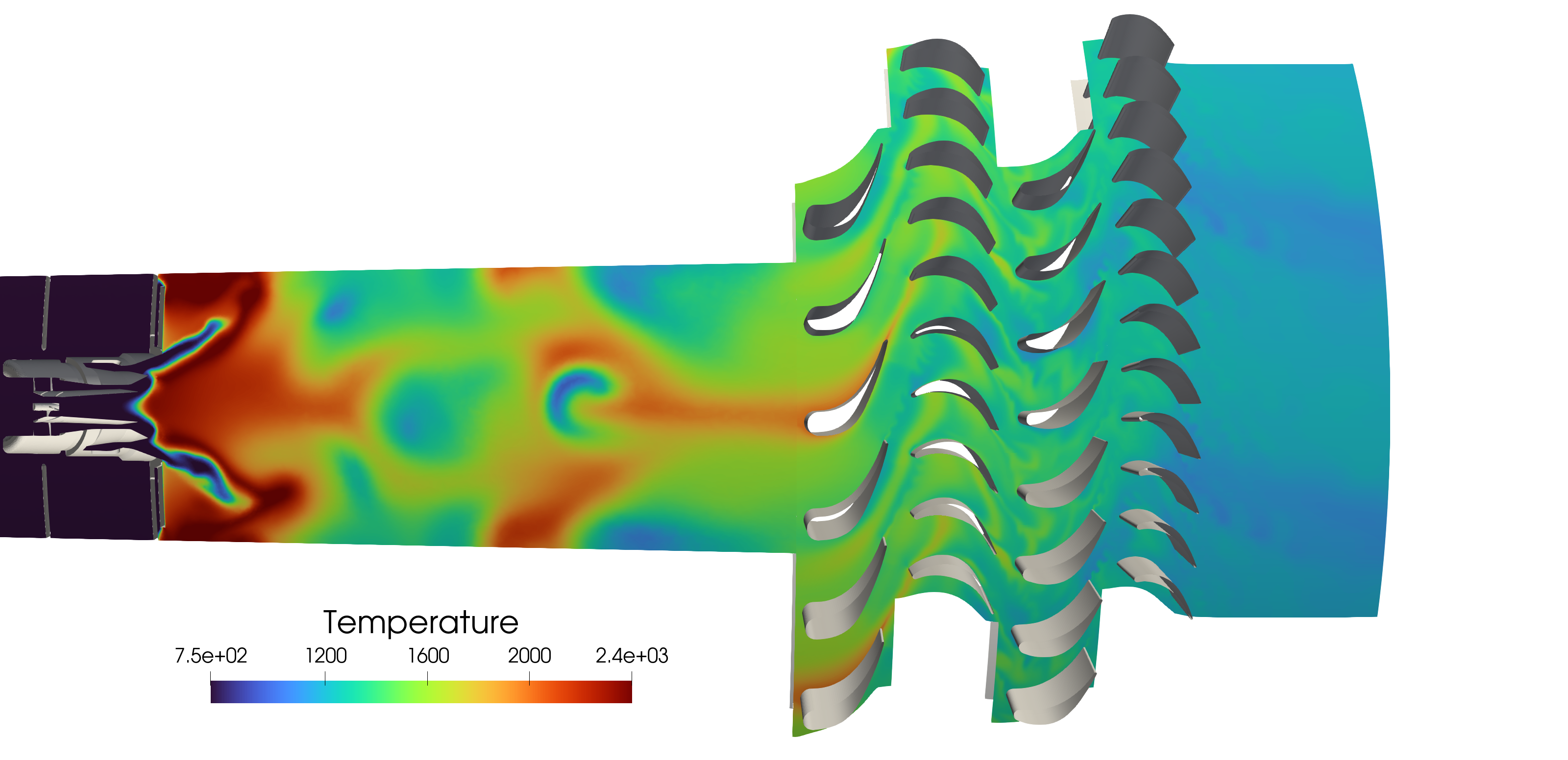}
    \begin{subfigure}[t]{0.84\linewidth}
        \centering
    \fbox{%
        \includegraphics[
            width=\linewidth,
            trim={0 420 352 530},
            clip
        ]{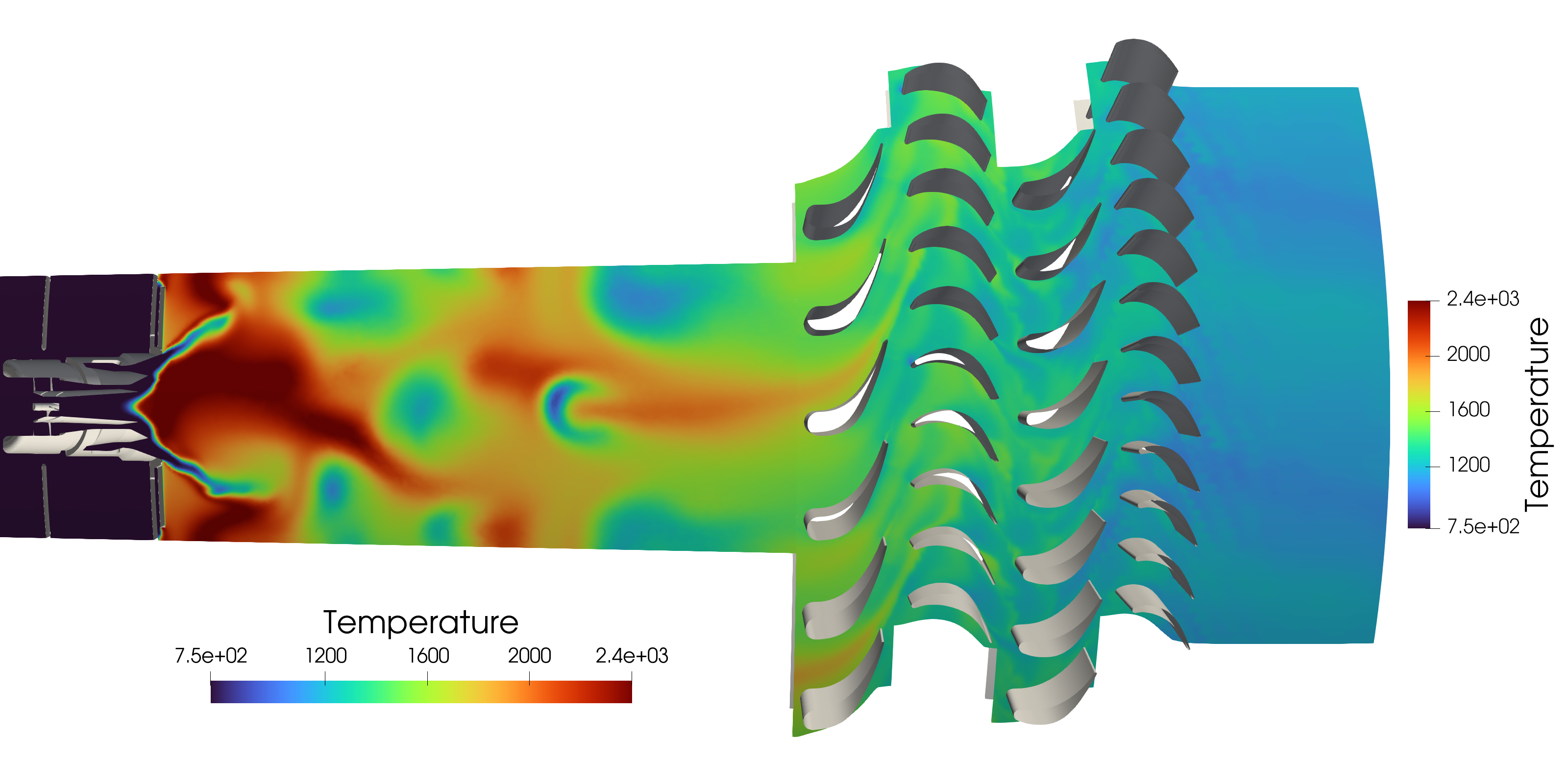}
        }
        \caption{$t/t_{rot} = 2.075$}
    \end{subfigure}
    \hfill
            \includegraphics[
            width=0.11\linewidth,
            trim={2868 440 20 550},
            clip
        ]{Figures-Results/react_full_view3_turbo_temp_0101.png}
    \begin{subfigure}[t]{0.84\linewidth}
        \centering
    \fbox{%
        \includegraphics[
            width=\linewidth,
            trim={0 420 352 530},
            clip
        ]{Figures-Results/react_full_view3_turbo_temp_0101.png}
        }
        \caption{$t/t_{rot} = 2.525$}
    \end{subfigure}
    \hfill
   \includegraphics[
            width=0.11\linewidth,
            trim={2868 440 20 550},
            clip
        ]{Figures-Results/react_full_view3_turbo_temp_0083.png}
    \caption{Snapshots of combustor-turbine temperature field}
    \label{fig:full-view}
\end{figure*}

\label{subsec:results-reacting}
Figure \ref{fig:full-view} shows three snapshots of the mid-span temperature field across the coupled combustor–turbine domain. In the combustor, the characteristic cross-sectional pattern generated by cooling and dilution holes is visible, together with the hot streak emerging between the dilution jets. This streak impinges directly onto the inlet guide vane at approximately the same circumferential position as the burner. As the flow evolves in time, the high-temperature region alternately impacts the pressure and suction sides of the IGV. Downstream of the vane, the streak is stretched and reshaped as it is convected through the first rotor row, giving rise to distinct pockets of elevated temperature. These are shed at rotor exit and subsequently interact with the second vane row, producing a periodically varying thermal loading pattern that remains clearly discernible up to the turbine outflow.

\begin{figure}
    \centering
        \includegraphics[width=\linewidth]{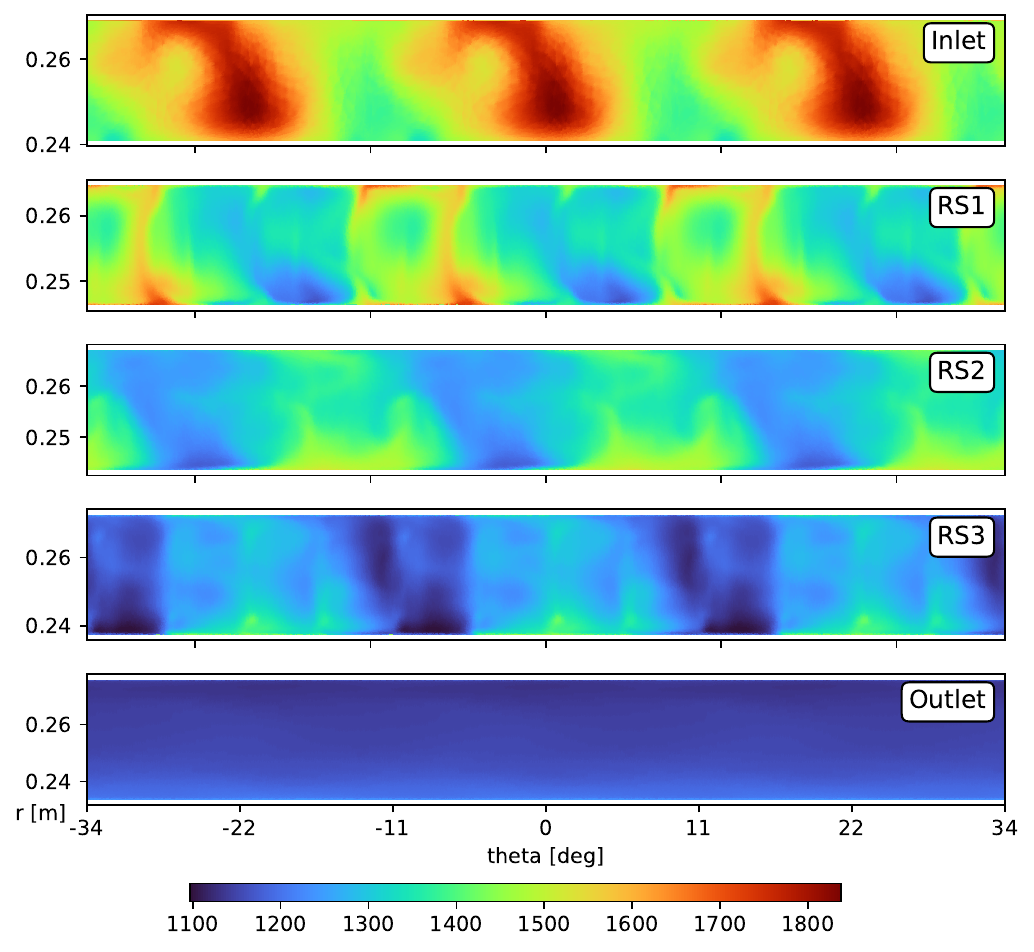}
    \caption{Favre-averaged temperature field at turbine inlet, rotor/stator interfaces and outlet}
    \label{fig:rotor-stators-tavg}
\end{figure}

Figure \ref{fig:rotor-stators-tavg} shows the Favre-averaged temperature field at the turbine inlet, rotor–stator interfaces, and turbine outlet. At machines' inlet, the characteristic hot temperature spot in the middle of a $20^{\circ}$ sector, also shown in the snapshot of 
Fig.~\ref{fig:combustor-exit},
remains clearly identifiable in the mean field, confirming the persistence of the burner-aligned hot region. In this configuration, the footprint of a dilution jet is also observed, partially mitigating the peak temperature, and a tendency towards higher temperatures near the outer casing is visible.

At the first rotor–stator interface, the hot spot is split between the pressure and suction sides of each vane, and the characteristic pattern of vane wakes are identifiable. The interaction with secondary flows, in particular the corner vortex, leads to an accumulation of hot fluid in the near-wall regions, with a concentration more pronounced in the near-hub region. Similar features appear at the second and third rotor–stator interfaces, although a progressive smoothing of the temperature gradients is evident. The second interface lies immediately downstream of a rotating row, so blade wakes averaged out, whereas they can be distinguished at the third interface.
Beyond the progressive smoothing observed, the hot streak exhibits a clear pattern of circumferential migration and deformation. As it is convected through the vane and rotor passages, the streak gradually shifts its azimuthal position and undergoes significant stretching, with the displacement being more pronounced toward the shroud. An angular shift of approximately $\approx 30^{\circ}$ is observed. This behaviour aligns with the trends identified in our earlier work~\cite{lopresti2022numerical}, considering the lower stage count of this configuration, and is consistent with the observations 
of~\cite{Adamczuk2012} and~\cite{Chi2019}.

\begin{figure}
    \centering
        \includegraphics[width=\linewidth]
        {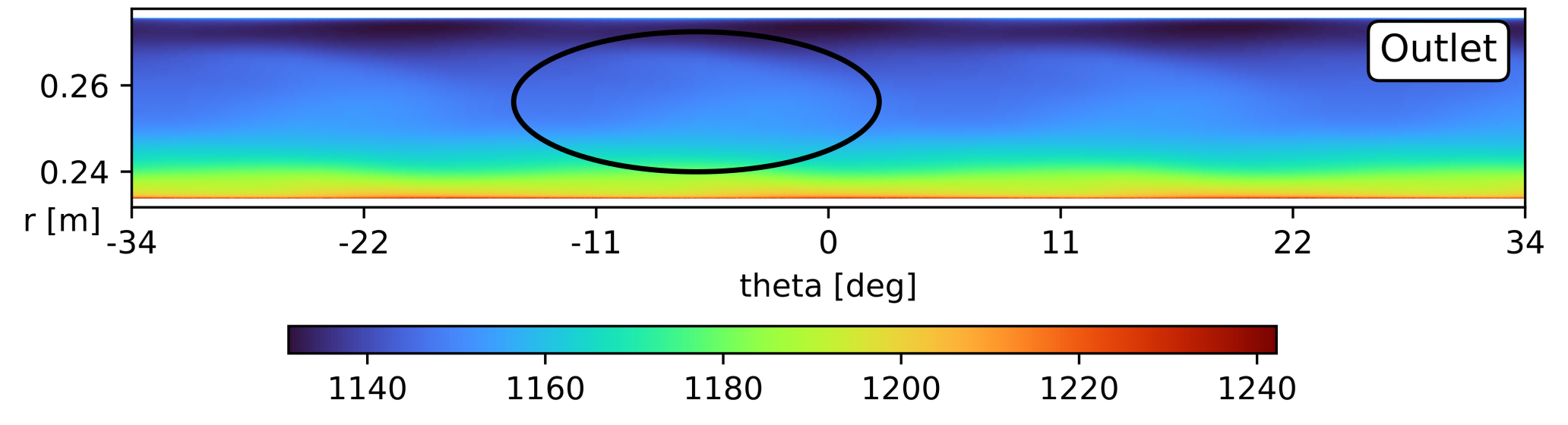}
    \caption{Favre-averaged temperature field at turbine outlet, rescaled to local min and max}
    \label{fig:rotor-stators-tavg-out-rescaled}
\end{figure}

At the turbine outlet, the mean temperature field appears nearly uniform; however, when plotted with an adapted colour scale (see Fig.~\ref{fig:rotor-stators-tavg-out-rescaled}), the residual footprint of the hot streak is still visible, consistent with the behaviour observed in the instantaneous fields of Fig.~\ref{fig:full-view}. 

\begin{figure}
    \centering
        \includegraphics[width=\linewidth]{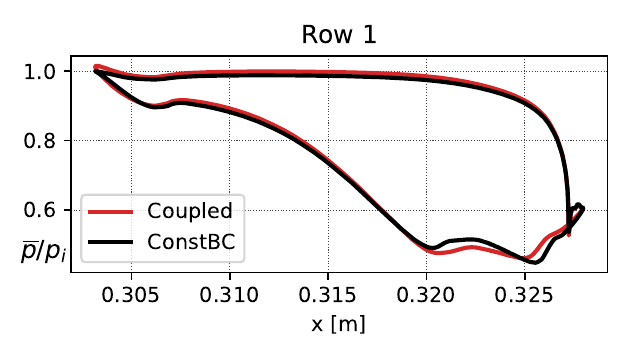}
    \caption{Normalized averaged pressure at midspan of IGVs: comparison of coupled vs standalone simulation with steady uniform inlet}
    \label{fig:profiles-p}
\end{figure}

Finally, results are analysed in terms of quantities directly linked to blade performance and durability, namely the flow field on the airfoil surfaces.
Figure~\ref{fig:profiles-p} compares the time averaged pressure distribution on the IGVs obtained from the standalone turbine simulation and from the coupled combustor–turbine computation. The two profiles show an excellent agreement over the entire chord and are practically identical for all IGVs. A similar level of agreement is also found in the downstream rows, not shown here for brevity. The analysis confirms that key performance metrics such as aerodynamic loading, flow turning, and work output can be reliably predicted by the standalone configuration, while the coupled setup doesn't introduce deviations in the mean pressure loading.

\begin{figure}
    \centering
        \includegraphics[width=\linewidth]{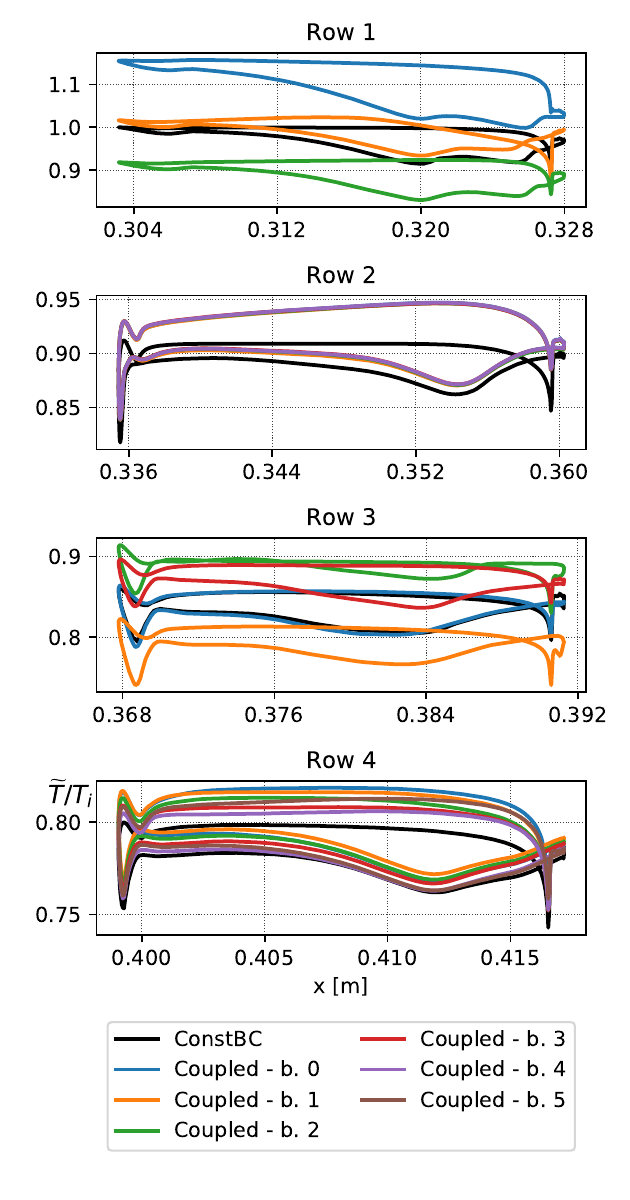}
    \caption{Normalized Favre-averaged temperature at midspan: comparison of coupled vs standalone simulation with steady uniform inlet}
    \label{fig:profiles-t}
\end{figure}

The same considerations do not hold for the Favre-averaged gas temperature profiles, shown in Fig.~\ref{fig:profiles-t} for all airfoils within each of the four rows.
Starting from row 1, and consistent with the instantaneous fields of Fig.~\ref{fig:full-view}, the IGV directly impacted by the hot streak exhibits markedly higher temperatures than in the standalone simulation. A second IGV, located in a colder inflow region, shows significantly lower temperatures, while an intermediate vane—intersected by a partially mixed stream—displays values comparable to those obtained with the uniform inlet condition.
Similar behaviour is observed for the vanes in row 3, where the thermal loading varies substantially with circumferential position, reflecting the non-uniform temperature distribution convected from upstream. The rotating rows, by contrast, experience a circumferentially averaged inflow; nevertheless, their surface temperatures remain systematically higher than those predicted by the standalone simulation with uniform inlet conditions.
The marked dependence of the thermal loading on the relative alignment between combustor features and turbine blade rows opens questions regarding clocking effects, suggesting that an optimized phasing between burners and airfoils could be exploited to mitigate peak thermal loads and potentially improve hot-section design.

\section{Conclusions and outlook}
This work presented a framework for the full, real-time coupling of a pressure-based solver with a density-based solver, applied in a Combustor/Turbine Interaction study. 
A key aspect of the methodology is the targeted treatment of chemistry at the interface, where finite-rate chemistry in the combustor is consistently coupled to an energy-augmented tabulated chemistry approach in the turbine, to account for the non-negligible compressibility effects of the expanding flow.
The coupled reacting simulation was demonstrated on the MYTHOS Virtual Test Rig.

The study showed modifications to the mean flow field in the final part of the combustor, most notably in the mean velocity field, which is directly influenced by the change in outflow conditions—driven primarily by the altered pressure boundary imposed by the coupled turbine domain.

In the turbine, the coupled simulation highlighted clear combustor-driven effects. Temperature and composition non-uniformities generated in the combustor persist across multiple blade rows: the hot streak is convected, stretched, and split by vane and rotor secondary flows, and its thermal imprint remains detectable even at the turbine exit despite substantial mixing. Aerodynamic performance metrics such as pressure loading remain essentially unchanged relative to standalone predictions, while substantial deviations are observed in the gas-temperature profiles, confirming the enhanced physical realism of the coupled approach. The pronounced circumferential variation in blade thermal loading, and its sensitivity to the relative alignment between combustor features and turbine airfoils, highlights the relevance of clocking effects and suggests potential benefits from optimized burner–airfoil phasing.

After this first demonstration of the method, the framework will be applied to Sustainable Aviation Fuel combustion and extended to include cooling flows, thereby allowing a realistic assessment of cooling effectiveness, which is strongly influenced by the unsteady inlet conditions. In addition, the modelling fidelity will be increased by adopting LES in the combustor and DES in the turbine, enabling a more accurate representation of unsteady mixing, secondary-flow dynamics, and thermal loading.

\section*{Acknowledgments}
This paper is supported by the European Union’s Horizon Europe research and innovation programme under Grant Agreement No. 101096286, project MYTHOS (Medium-range hybrid low-pollution flexi-fuel/hydrogen sustainable engine). Views and opinions expressed are, however, those of the authors only and do not necessarily reflect those of the European Union or CINEA. Neither the European Union nor the granting authority can be held responsible for them.

We acknowledge EuroHPC Joint Undertaking for awarding us access to LUMI at CSC, Finland.

\bibliographystyle{unsrtnat}
\bibliography{references}

@article{butler1989redistribution,
  author    = {T. L. Butler and O. P. Sharma and H. D. Joslyn and R. P. Dring},
  title     = {Redistribution of an inlet temperature distortion in an axial flow turbine stage},
  journal   = {J. Propul. Power},
  volume    = {5},
  number    = {1},
  pages     = {64-71},
  year      = {1989},
  doi       = {10.2514/3.23116}
}

@article{jenny2012hot,
  author    = {P. Jenny and C. Lenherr and R. S. Abhari and A. Kalfas},
  title     = {Effect of hot streak migration on unsteady blade row interaction in an axial turbine},
  journal   = {ASME J. Turbomach.},
  volume    = {134},
  number    = {5},
  pages     = {051020},
  year      = {2012},
  doi       = {10.1115/1.4004447}
}

@article{povey2005hot,
  author    = {T. Povey and K. S. Chana and T. V. Jones and J. Hurrion},
  title     = {The Effect of Hot-Streaks on HP Vane Surface and Endwall Heat Transfer: An Experimental and Numerical Study},
  journal   = {J. Turbomach.},
  volume    = {129},
  number    = {1},
  pages     = {32-43},
  year      = {2005},
  doi       = {10.1115/1.2370748}
}

@article{rai1990navier,
  author    = {M. M. Rai and R. P. Dring},
  title     = {Navier-Stokes analyses of the redistribution of inlet temperature distortions in a turbine},
  journal   = {J. Propul. Power},
  volume    = {6},
  number    = {3},
  pages     = {276-282},
  year      = {1990},
  doi       = {10.1115/1.4002039}
}

@article{dorney1992hot,
  author    = {D. J. Dorney and R. L. Davis and D. E. Edwards and N. K. Madavan},
  title     = {Unsteady analysis of hot streak migration in a turbine stage},
  journal   = {J. Propul. Power},
  volume    = {8},
  number    = {2},
  pages     = {520-529},
  year      = {1992},
  doi       = {10.2514/3.23507}
}

@inproceedings{takahashi1990unsteady,
  author    = {R. Takahashi and R. Ni},
  title     = {Unsteady Euler analysis of the redistribution of an inlet temperature distortion in a turbine},
  booktitle = {Proceedings of the 26th Joint Propulsion Conference},
  address   = {Orlando, FL},
  year      = {1990},
  doi       = {10.2514/6.1990-2262}
}

@inproceedings{ong2008hot,
  author    = {J. Ong and R. J. Miller},
  title     = {Hot streak and vane coolant migration in a downstream rotor},
  booktitle = {Proceedings of ASME Turbo Expo 2008},
  address   = {Berlin, Germany},
  year      = {2008},
  doi       = {10.1115/GT2008-50971}
}

@phdthesis{legrenzi2017coupled,
  title={A Coupled CFD Approach for Combustor-Turbine Interaction},
  author={Legrenzi, Paolo},
  school={Loughborough University},
  year={2017},
  note={Available at: \url{https://repository.lboro.ac.uk/articles/thesis/A_coupled_CFD_approach_for_combustor-turbine_interaction/9214682}}
}

@inproceedings{schlueter2004,
    author    = {Schl{\"u}ter, J. U. and Kim, S. and Wu, X. and Alonso, J. J. and Pitsch, H.},
    title     = {Large-Scale Integrated LES--RANS Simulations of a Gas Turbine Engine},
    booktitle = {Center for Turbulence Research, Annual Research Briefs},
    pages     = {105--116},
    address   = {Stanford University / NASA Ames},
    publisher = {Center for Turbulence Research},
    year      = {2004}
}

@inproceedings{schlueter2005,
    author    = {Schl{\"u}ter, J. U. and Wu, X. and van der Weide, E. and Hahn, S. and Alonso, J. J. and Pitsch, H.},
    title     = {Multi-Code Simulations: A Generalized Coupling Approach},
    booktitle = {17th AIAA Computational Fluid Dynamics Conference},
    year      = {2005},
    doi       = {10.2514/6.2005-4997}
}

@inproceedings{Turner2002,
    author    = {Turner, Mark and Ryder, Rob and Celestina, Mark and Moder, Jeff and Liu, Nan-Suey and Adamczyk, John and Veres, Joe},
    title     = {High Fidelity 3D Turbofan Engine Simulation with Emphasis on Turbomachinery--Combustor Coupling},
    booktitle = {38th AIAA/ASME/SAE/ASEE Joint Propulsion Conference \& Exhibit},
    year      = {2002},
    doi       = {10.2514/6.2002-3769}
}

@inproceedings{Turner2003,
    author    = {Turner, Mark and Norris, Andrew and Veres, Joe},
    title     = {High Fidelity 3D Simulation of the GE90 (Invited)},
    booktitle = {33rd AIAA Fluid Dynamics Conference and Exhibit},
    year      = {2003},
    doi       = {10.2514/6.2003-3996}
}

@inproceedings{Turner2010,
    author    = {Turner, Mark},
    title     = {Lessons Learned from the GE90 3-D Full Engine Simulations},
    booktitle = {48th AIAA Aerospace Sciences Meeting Including the New Horizons Forum and Aerospace Exposition},
    year      = {2010},
    doi       = {10.2514/6.2010-1606}
}

@article{PerezArroyo2021a,
    author    = {P{\'e}rez Arroyo, Carlos and Dombard, J{\'e}r{\^o}me and Duchaine, Florent and Gicquel, Laurent and Martin, Benjamin and Odier, Nicolas and Staffelbach, Gabriel},
    title     = {Towards the Large-Eddy Simulation of a Full Engine: Integration of a 360 Azimuthal Degrees Fan, Compressor and Combustion Chamber. Part I: Methodology and Initialisation},
    journal   = {Journal of the Global Power and Propulsion Society},
    year      = {2021},
    number    = {May},
    pages     = {1--16},
    doi       = {10.33737/jgpps/133115}
}

@article{PerezArroyo2021b,
    author    = {P{\'e}rez Arroyo, Carlos and Dombard, J{\'e}r{\^o}me and Duchaine, Florent and Gicquel, Laurent and Martin, Benjamin and Odier, Nicolas and Staffelbach, Gabriel},
    title     = {Towards the Large-Eddy Simulation of a Full Engine: Integration of a 360 Azimuthal Degrees Fan, Compressor and Combustion Chamber. Part II: Comparison Against Stand-Alone Simulations},
    journal   = {Journal of the Global Power and Propulsion Society},
    year      = {2021},
    number    = {May},
    pages     = {1--16},
    doi       = {10.33737/jgpps/133116},
}

@article{Miki2023,
    author    = {Miki, Kenji and Wey, Thomas and Moder, Jeffrey},
    title     = {Computational Study on Fully Coupled Combustor--Turbine Interactions},
    journal   = {Journal of Propulsion and Power},
    volume    = {39},
    number    = {4},
    pages     = {540--553},
    year      = {2023},
    doi       = {10.2514/1.B38501}
}

@inproceedings{Miki2025,
    author    = {Miki, Kenji and Wey, Changju T. and Turner, Mark G. and Moder, Jeffrey P.},
    title     = {Numerical Investigation of Combustor--Turbine Interactions With Cooling Air Flows Included in High-Pressure Turbine},
    booktitle = {AIAA SCITECH 2025 Forum},
    year      = {2025},
    doi       = {10.2514/6.2025-2089}
}

@inproceedings{salvadori2012,
  author    = {S. Salvadori and G. Riccio and M. Insinna and F. Martelli},
  title     = {Analysis of combustor/vane interaction with decoupled and loosely coupled approaches},
  booktitle = {Proceedings of ASME Turbo Expo 2012},
  address   = {Copenhagen, Denmark},
  year      = {2012},
  doi       = {10.1115/GT2012-69038}
}

@phdthesis{papadogiannis2015,
    author    = {Papadogiannis, Dimitrios},
    title     = {Coupled Large Eddy Simulations of Combustion Chamber--Turbine Interactions},
    school    = {Institut National Polytechnique de Toulouse (INPT)},
    year      = {2015},
    type      = {Ph.D. dissertation},
    url       = {https://oatao.univ-toulouse.fr/14169/}
}

@article{Eurocontrol2021SAF,
  author       = {EUROCONTROL},
  title        = {Can sustainable aviation fuels help us decarbonise aviation?},
  journal      = {EUROCONTROL Aviation Brief},
  year         = {2021},
  url          = {https://www.eurocontrol.int/article/can-sustainable-aviation-fuels-help-us-decarbonise-aviation}
}

@article{Voigt2021,
    author  = {Voigt, Christian and Kleine, Jannis and Sauer, Daniel and Moore, Richard H. and Bräuer, Tobias and Le Clercq, Philippe and Kaufmann, Stephan and Scheibe, Martin and Jurkat-Witschas, Tina and Aigner, Manfred and Bauder, Ulrich and Boose, Yvonne and Borrmann, Stephan and Crosbie, Eric C. and Diskin, Glenn S. and DiGangi, Joshua P. and Hahn, Verena and Heckl, Caroline and Huber, Felix and Nowak, John B. and Rapp, Markus and Rauch, Benedikt and Robinson, Charles E. and Schripp, Tobias and Shook, Melissa A. and Winstead, Elizabeth L. and Ziemba, Luke D. and Schlager, Hans and Anderson, Bruce E.},
    title   = {Cleaner burning aviation fuels can reduce contrail cloudiness},
    journal = {Communications Earth \& Environment},
    volume  = {2},
    pages   = {114},
    year    = {2021},
    doi     = {10.1038/s43247-021-00174-y}
}

@article{Lee2021,
    author  = {Lee, D. S. and Fahey, D. W. and Skowron, A. and Allen, M. R. and Burkhardt, U. and Chen, Q. and Doherty, S. J. and Freeman, S. and Forster, P. M. and Fuglestvedt, J. and Gettelman, A. and De León, R. R. and Lim, L. L. and Lund, M. T. and Millar, R. J. and Owen, B. and Penner, J. E. and Pitari, G. and Prather, M. J. and Sausen, R. and Wilcox, L. J.},
    title   = {The Contribution of Global Aviation to Anthropogenic Climate Forcing for 2000 to 2018},
    journal = {Atmospheric Environment},
    volume  = {244},
    pages   = {117834},
    year    = {2021},
    issn    = {1352-2310},
    doi     = {10.1016/j.atmosenv.2020.117834}
}

@article{Moore2015,
    author  = {Moore, Richard H. and Shook, Michael and Beyersdorf, Andreas and Corr, Chelsea and Herndon, Scott and Knighton, W. Berk and Miake-Lye, Richard and Thornhill, K. Lee and Winstead, Edward L. and Yu, Zhenhong and Ziemba, Luke D. and Anderson, Bruce E.},
    title   = {Influence of Jet Fuel Composition on Aircraft Engine Emissions: A Synthesis of Aerosol Emissions Data from the NASA APEX, AAFEX, and ACCESS Missions},
    journal = {Energy \& Fuels},
    volume  = {29},
    number  = {4},
    pages   = {2591--2600},
    year    = {2015},
    doi     = {10.1021/ef502618w},
    issn    = {0887-0624},
    publisher = {American Chemical Society}
}

@article{Kang2019a,
    author    = {Kang, Dongil and Kim, Doohyun and Kalaskar, Vickey and Violi, Angela and Boehman, André L.},
    title     = {Experimental Characterization of Jet Fuels under Engine Relevant Conditions -- Part 1: Effect of Chemical Composition on Autoignition of Conventional and Alternative Jet Fuels},
    journal   = {Fuel},
    volume    = {239},
    pages     = {1388--1404},
    year      = {2019},
    doi       = {10.1016/j.fuel.2018.10.005},
    issn      = {0016-2361}
}

@article{Peiffer2019,
    author    = {Peiffer, Erin E. and Heyne, Joshua S. and Colket, Meredith},
    title     = {Sustainable Aviation Fuels Approval Streamlining: Auxiliary Power Unit Lean Blowout Testing},
    journal   = {AIAA Journal},
    volume    = {57},
    number    = {11},
    pages     = {4854--4862},
    year      = {2019},
    doi       = {10.2514/1.J058348},
    issn      = {0001-1452}
}

@techreport{HolladayAbdullahHeyne2020,
  author       = {Holladay, Johnathan and Abdullah, Zia and Heyne, Joshua},
  title        = {Sustainable Aviation Fuel: Review of Technical Pathways},
  institution  = {U.S. Department of Energy},
  number       = {DOE/EE-2041},
  year         = {2020},
  url          = {https://www.energy.gov/sites/prod/files/2020/09/f78/beto-sust-aviation-fuel-sep-2020.pdf}
}

@article{LinNurazaqWang2023,
  author       = {Lin, Jhe-Kai and Nurazaq, Warit Abi and Wang, Wei-Cheng},
  title        = {The properties of sustainable aviation fuel I: Spray characteristics},
  journal      = {Energy},
  volume       = {283},
  pages        = {129125},
  year         = {2023},
  doi          = {10.1016/j.energy.2023.129125}
}

@article{HarlassDischlKaufmann2024,
  author       = {Harlass, Theresa and Dischl, Rebecca Katharina and Kaufmann, Stefan and Märkl, Raphael Satoru and Sauer, Daniel and Scheibe, Monika and Stock, Paul and Bräuer, Tiziana and Dörnbrack, Andreas and Roiger, Anke-Elisabeth and Schlager, Hans and Schumann, Ulrich and Pühl, Magdalena and Schripp, Tobias and Grein, Tobias and Bondorf, Linda and Renard, Charles and Gauthier, Maxime and Johnson, Mark and Luff, Darren and Madden, Paul and Swann, Peter and Ahrens, Denise and Sallinen, Reetu and Voigt, Christiane},
  title        = {Measurement report: In-flight and ground-based measurements of nitrogen oxide emissions from latest-generation jet engines and 100 \% sustainable aviation fuel},
  journal      = {Atmospheric Chemistry and Physics},
  volume       = {24},
  pages        = {11807--11822},
  year         = {2024},
  doi          = {10.5194/acp-24-11807-2024}
}

@inproceedings{Cha2012,
  author    = {C. M. Cha and S. Hong and P. T. Ireland and P. Denman and V. Savarianandam},
  title     = {Experimental and Numerical Investigation of Combustor-Turbine Interaction Using an Isothermal, Nonreacting Tracer},
  booktitle = {Proceedings of the ASME Turbo Expo},
  year      = {2012}
}

@article{Jacobi2017,
  author  = {S. Jacobi and B. Rosic},
  title   = {Thermal Investigation of Integrated Combustor Vane Concept under Engine-Realistic Conditions},
  journal = {Journal of Turbomachinery},
  volume  = {139},
  number  = {2},
  pages   = {021005},
  year    = {2017},
  doi     = {10.1115/1.4034433}
}

@inproceedings{Govert2019,
  author    = {S. Gövert and F. Ferraro and A. Krumme and C. Buske and M. Tegeler and F. Kocian and F. {di Mare}},
  title     = {Investigation of Combustor-Turbine-Interaction in a Rotating Cooled Transonic High-Pressure Turbine Test Rig: Part 2---Numerical Modelling and Simulation},
  booktitle = {ASME Turbo Expo: Power for Land, Sea, and Air},
  volume    = {58561},
  year      = {2019},
  pages     = {V02BT42A005},
  doi       = {10.1115/GT2019-90439}
}

@article{Klapdor2013,
  author  = {E. V. Klapdor and F. {di Mare} and W. Kollmann and J. Janicka},
  title   = {A Compressible Pressure-Based Solution Algorithm for Gas Turbine Combustion Chambers Using the PDF/FGM Model},
  journal = {Flow, Turbulence and Combustion},
  volume  = {91},
  pages   = {209--247},
  year    = {2013},
  doi     = {10.1007/s10494-013-9495-6}
}

@article{Tomasello2023,
  author  = {S. G. Tomasello and R. Meloni and L. Andrei and A. Andreini},
  title   = {Study of Combustor--Turbine Interactions by Performing Coupled and Decoupled Hybrid RANS-LES Simulations under Representative Engine-like Conditions},
  journal = {Energies},
  volume  = {16},
  number  = {14},
  pages   = {5395},
  year    = {2023},
  doi     = {10.3390/en16145395}
}

@inproceedings{LoPresti2025,
  author    = {Federico {Lo Presti} and Francesca {di Mare}},
  title     = {All Mach flows density-based / pressure-based coupling: method and preliminary results on the example of a rim seal flow in a one-stage axial turbine},
  booktitle = {16th European Conference on Turbomachinery Fluid dynamics \& Thermodynamics (ETC16)},
  year      = {2025},
  paperid   = {ETC2025-272},
  doi       = {10.29008/ETC2025-272}
}

@article{vicquelin2011coupling,
  title={Coupling tabulated chemistry with compressible CFD solvers},
  author={Vicquelin, Ronan and Fiorina, Benoit and Payet, Sandra and Darabiha, Nasser and Gicquel, Olivier},
  journal={Proceedings of the Combustion Institute},
  volume={33},
  number={1},
  pages={1481--1488},
  year={2011},
  publisher={Elsevier}
}

@article{terrapon2009flamelet,
  title={A flamelet-based model for supersonic combustion},
  author={Terrapon, Vincent and Ham, Frank and Pecnik, Rene and Pitsch, Heinz},
  journal={Annual research briefs},
  year={2009},
  publisher={Center for Turbulence Research, National Aeronautics and Space~…}
}

@article{mittal2013flamelet,
  title={A flamelet model for premixed combustion under variable pressure conditions},
  author={Mittal, Varun and Pitsch, Heinz},
  journal={Proceedings of the Combustion Institute},
  volume={34},
  number={2},
  pages={2995--3003},
  year={2013},
  publisher={Elsevier}
}

@article{saghafian2015efficient,
  title={An efficient flamelet-based combustion model for compressible flows},
  author={Saghafian, Amirreza and Terrapon, Vincent E and Pitsch, Heinz},
  journal={Combustion and Flame},
  volume={162},
  number={3},
  pages={652--667},
  year={2015},
  publisher={Elsevier}
}

@inproceedings{ma2017numerical,
  title={Numerical framework for transcritical real-fluid reacting flow simulations using the flamelet progress variable approach},
  author={Ma, Peter C and Banuti, Daniel and Hickey, Jean-Pierre and Ihme, Matthias},
  booktitle={55th AIAA Aerospace Sciences Meeting},
  pages={0143},
  year={2017}
}

@article{fiorina2003modelling,
  title={Modelling non-adiabatic partially premixed flames using flame-prolongation of ILDM},
  author={Fiorina, Benoit and Baron, Romain and Gicquel, Olivier and Thevenin, D and Carpentier, St{\'e}phane and Darabiha, Nasser},
  journal={Combustion Theory and Modelling},
  volume={7},
  number={3},
  pages={449},
  year={2003},
  publisher={IOP Publishing}
}

@inproceedings{donini2013numerical,
  title={Numerical simulations of a premixed turbulent confined jet flame using the flamelet generated manifold approach with heat loss inclusion},
  author={Donini, A and Martin, SM and Bastiaans, RJM and van Oijen, JA and De Goey, LPH},
  booktitle={Turbo Expo: Power for Land, Sea, and Air},
  volume={55102},
  pages={V01AT04A024},
  year={2013},
  organization={American Society of Mechanical Engineers}
}

@article{ketelheun2013heat,
  title={Heat transfer modeling in the context of large eddy simulation of premixed combustion with tabulated chemistry},
  author={Ketelheun, Anja and Kuenne, Guido and Janicka, Johannes},
  journal={Flow, turbulence and combustion},
  volume={91},
  number={4},
  pages={867--893},
  year={2013},
  publisher={Springer}
}

@article{proch2015modeling,
  title={Modeling heat loss effects in the large eddy simulation of a model gas turbine combustor with premixed flamelet generated manifolds},
  author={Proch, F and Kempf, AM},
  journal={Proceedings of the combustion institute},
  volume={35},
  number={3},
  pages={3337--3345},
  year={2015},
  publisher={Elsevier}
}

@article{ihme2008modeling,
  title={Modeling of radiation and nitric oxide formation in turbulent nonpremixed flames using a flamelet/progress variable formulation},
  author={Ihme, Matthias and Pitsch, Heinz},
  journal={Physics of Fluids},
  volume={20},
  number={5},
  year={2008},
  publisher={AIP Publishing}
}

@techreport{donndorf2025baseline,
  author       = {Jan Donndorf and Pierre Vauquelin and Francesca {di Mare} and Xue Song Bai and Christer Fureby},
  title        = {Virtual twin of baseline combustor},
  institution  = {European Commission, Horizon Europe Project MYTHOS},
  type         = {Technical Report},
  number       = {101096286.D3.2},
  year         = {2025},
  url          = {https://cordis.europa.eu/project/id/101096286/results},
}

@techreport{vauquelin2025optimal,
  author       = {Pierre Vauquelin and Jan Donndorf and Francesca {di Mare} and Xue Song Bai and Christer Fureby},
  title        = {Optimal enhanced combustor design for fuel flexibility},
  institution  = {European Commission, Horizon Europe Project MYTHOS},
  type         = {Technical Report},
  number       = {101096286.D3.3},
  year         = {2025},
  url          = {https://cordis.europa.eu/project/id/101096286/results},
}

@inproceedings{vauquelin2026design,
  author       = {Pierre Vauquelin and Jan Donndorf and Federico {Lo Presti} and Francesca {di Mare} and Xue-Song Bai and Christer Fureby},
  title        = {Design, Simulation, and Virtual Certification Assessment of a medium range Aeroengine Combustor},
 booktitle = {AIAA SCITECH 2026 Forum},
    year      = {2025},
}

@inproceedings{Donndorf2025,
  author    = {J. Donndorf and {Lo Presti}, F. and {di Mare}, F.},
  title     = {Thermodynamic Design of an Aeroengine for Sustainable Aviation Fuels with a Focus on 3D Components Conception - High Pressure Turbine},
  booktitle = {Proceedings of the 16th European Turbomachinery Conference (ETC16)},
  year      = {2025},
  address   = {Hannover, Germany},
  month     = {March 24-28}
}

@techreport{lopresti2025virtual,
  author       = {Federico {Lo Presti} and Nunzio Natale and Francesca {di Mare}},
  title        = {Virtual twin of expansion and compression section},
  institution  = {European Commission, Horizon Europe Project MYTHOS},
  type         = {Technical Report},
  number       = {101096286.D3.4},
  year         = {2025},
  url          = {https://cordis.europa.eu/project/id/101096286/results},
}

@article{post2021a,
  title = {Large {{Eddy Simulation}} of a {{Condensing Wet Steam Turbine Cascade}}},
  author = {Post, Pascal and Winhart, Benjamin and {di Mare}, Francesca},
  date = {2021-02-01},
  journaltitle = {Journal of Engineering for Gas Turbines and Power},
  volume = {143},
  number = {2},
  pages = {021016},
  issn = {0742-4795, 1528-8919},
  doi = {10.1115/1.4049348}
}

@article{ziaja2021numerical,
  title={Numerical Investigation of a Partially Loaded Supersonic Organic Rankine Cycle Turbine Stage},
  author={Ziaja, Karl and Post, Pascal and Sembritzky, Marwick and Schramm, Andreas and Willers, Ole and Kunte, Harald and Seume, Joerg R and {di Mare}, Francesca},
  journal={Journal of Engineering for Gas Turbines and Power},
  volume={143},
  number={6},
  pages={061014},
  year={2021},
  publisher={American Society of Mechanical Engineers}
}

@article{lopresti2022numerical,
  title={Numerical investigation of unsteady combustor turbine interaction for flexible power generation},
  author={Lo Presti, Federico and Sembritzky, Marwick and Winhart, Benjamin and Post, Pascal and {di Mare}, Francesca and Wiedermann, Alexander and Greving, Johannes and Krewinkel, Robert},
  journal={Journal of Turbomachinery},
  volume={144},
  number={2},
  pages={021003},
  year={2022},
  publisher={American Society of Mechanical Engineers}
}

@article{karaefe2021,
  title = {Numerical Analysis of a Centrifugal Compressor Operating with Supercritical {{CO2}}},
  author = {Karaefe, Renan and Post, Pascal and Sembritzky, Marwick and Schramm, Andreas and {di Mare}, Francesca and Kunick, Matthias and Gampe, Uwe},
  namea = {DuEPublico: Duisburg-Essen Publications Online, University Of Duisburg-Essen},
  nameatype = {collaborator},
  date = {2021-03-30},
  journaltitle = {Conference Proceedings of the European sCO2 Conference4th European sCO2 Conference for Energy Systems: March 23-24},
  volume = {2021},
  pages = {p. 230},
  publisher = {DuEPublico: Duisburg-Essen Publications online, University of Duisburg-Essen, Germany},
  doi = {10.17185/DUEPUBLICO/73966}
}

@article{Adamczuk2012,
    author    = {Adamczuk, Rafael R. and Seume, Joerg R.},
    title     = {Time Resolved Full-Annulus Computations of a Turbine with Inhomogeneous Inlet Conditions},
    journal   = {International Journal of Gas Turbine, Propulsion and Power Systems},
    volume    = {4},
    pages     = {1--7},
    year      = {2012},
    issn      = {1882-5098},
    publisher = {Gas Turbine Society of Japan},
    url       = {https://www.gtsj.or.jp/journal/ejournal/Vol.4/No.1/Vol.4%20No.1-1.pdf},
    note      = {Peer-reviewed}
}

@article{Chi2019,
    author    = {Chi, Zhongran and Liu, Haiqing and Zang, Shusheng and Pan, Chengxiong and Jiao, Guangyun},
    title     = {Full-Annulus URANS Study on the Transportation of Combustion Inhomogeneity in a Four-Stage Cooled Turbine},
    journal   = {ASME Journal of Turbomachinery},
    volume    = {141},
    number    = {11},
    pages     = {111003},
    year      = {2019},
    doi       = {10.1115/1.4044661},
    url       = {https://doi.org/10.1115/1.4044661},
    issn      = {0889-504X}
}

@article{weller1998tensorial,
  title        = {A tensorial approach to computational continuum mechanics using object oriented techniques},
  author       = {Weller, H. G. and Tabor, G. and Jasak, H. and Fureby, C.},
  journal      = {Computers in Physics},
  volume       = {12},
  number       = {6},
  pages        = {620--631},
  year         = {1998},
  doi          = {10.1063/1.168744}
}

@book{poinsot2005turbulentcombustion,
  title        = {Theoretical and Numerical Combustion},
  author       = {Poinsot, T. and Veynante, D.},
  year         = {2005},
  publisher    = {R. T. Edwards Inc.}
}

@incollection{menon2010computationalcombustion,
  title        = {Computational combustion},
  author       = {Menon, S. and Fureby, C.},
  booktitle    = {Encyclopedia of Aerospace Engineering},
  year         = {2010},
  publisher    = {John Wiley \& Sons, Ltd},
  doi          = {10.1002/9780470686652.eae063}
}

@article{sabelnikov2013lesmultiphase,
  title        = {{LES} combustion modeling for high {Re} flames using a multi-phase analogy},
  author       = {Sabelnikov, V. and Fureby, C.},
  journal      = {Combustion and Flame},
  volume       = {160},
  number       = {1},
  pages        = {83--96},
  year         = {2013},
  doi          = {10.1016/j.combustflame.2012.09.008}
}

@article{damkohler1936einfluesse,
  author      = {Damk{\"o}hler, Gerhard},
  title       = {Einfl{\"u}sse der Str{\"o}mung, Diffusion und des W{\"a}rme{\"u}berganges auf die Leistung von Reaktions{\"o}fen},
  journal     = {Zeitschrift f{\"u}r Elektrochemie},
  volume      = {42},
  number      = {12},
  pages       = {846--862},
  year        = {1936}
}

@article{dukowicz1980particlemodel,
  title        = {A particle-fluid numerical model for liquid sprays},
  author       = {Dukowicz, J. K.},
  journal      = {Journal of Computational Physics},
  volume       = {35},
  number       = {2},
  pages        = {229--253},
  year         = {1980},
  doi          = {10.1016/0021-9991(80)90087-X}
}

@article{apte2003atomizingspray,
  title        = {{LES} of atomizing spray with stochastic modeling of secondary breakup},
  author       = {Apte, S. V. and Gorokhovski, M. and Moin, P.},
  journal      = {International Journal of Multiphase Flow},
  volume       = {29},
  number       = {9},
  pages        = {1503--1522},
  year         = {2003},
  doi          = {10.1016/S0301-9322(03)00111-3}
}

@article{reitz1987atomizationmechanisms,
  title        = {Mechanisms of atomization processes in high-pressure vaporizing sprays},
  author       = {Reitz, R. D.},
  journal      = {Atomization and Spray Technology},
  volume       = {3},
  pages        = {309--337},
  year         = {1987}
}

@article{reitz1986dropbreakup,
  title        = {Effect of drop breakup on fuel sprays},
  author       = {Reitz, R. D. and Diwakar, R.},
  journal      = {SAE Transactions},
  volume       = {95},
  pages        = {218--227},
  year         = {1986}
}

@article{ranz1952evaporationdrops,
  title        = {Evaporation from drops},
  author       = {Ranz, W. E. and Marshall, W. R.},
  journal      = {Chemical Engineering Progress},
  volume       = {48},
  pages        = {141--146},
  year         = {1952}
}

@phdthesis{zettervall2021thesis,
  title        = {Methodology for developing reduced reaction mechanisms, and their use in combustion simulations},
  author       = {Zettervall, N.},
  school       = {Lund University},
  year         = {2021}
}

@article{AkerblomZettervallFureby,
  title        = {Comparing Chemical Reaction Mechanisms for Jet Fuel in Turbulent Premixed Combustion Simulations},
  author       = {{\AA}kerblom, A. and Zettervall, N. and Fureby, C.},
  journal      = {AIAA Journal},
  volume       = {63},
  number       = {9},
  pages        = {3493--4009},
  year         = {2025},
  doi          = {10.2514/1.J065162}
}

@article{yoshikawa2009noxmechanism,
  title        = {Development of an improved NO$_x$ reaction mechanism for low temperature diesel combustion modeling},
  author       = {Yoshikawa, T. and Reitz, R. D.},
  journal      = {SAE International Journal of Engines},
  volume       = {1},
  number       = {1},
  pages        = {1105--1117},
  year         = {2009}
}

@article{CRECK,
  title        = {New Reaction Classes in the Kinetic Modeling of Low Temperature Oxidation of n-Alkanes},
  author       = {Ranzi, E. and Cavallotti, C. and Cuoci, A. and Frassoldati, A. 
                  and Pelucchi, M. and Faravelli, T.},
  journal      = {Combustion and Flame},
  volume       = {162},
  number       = {5},
  pages        = {1679--1691},
  year         = {2015},
  doi          = {10.1016/j.combustflame.2014.11.030}
}

@article{HyChemJetA,
  title        = {HyChem Modeling of Combustion Kinetics of a Bio-Derived Jet Fuel and Its Blends with a Conventional Jet A},
  author       = {Wang, K. and Xu, R. and Parise, T. and Shao, J. and Movaghar, A. 
                  and Lee, D. J. and Park, J.-W. and Gao, Y. and Lu, T. 
                  and Egolfopoulos, F. N. and Davidson, D. F. and Hanson, R. K. 
                  and Bowman, C. T. and Wang, H.},
  journal      = {Combustion and Flame},
  volume       = {198},
  pages        = {477--489},
  year         = {2018},
  doi          = {10.1016/j.combustflame.2018.07.012}
}

@article{Hui2013,
  title        = {Laminar Flame Speeds of Transportation-Relevant Hydrocarbons and Jet Fuels at Elevated Temperatures and Pressures},
  author       = {Hui, X. and Sung, C.-J.},
  journal      = {Fuel},
  volume       = {109},
  pages        = {191--200},
  year         = {2013},
  doi          = {10.1016/j.fuel.2012.12.084}
}

@article{Kumar2011,
  title        = {Laminar Flame Speeds and Extinction Limits of Conventional and Alternative Jet Fuels},
  author       = {Kumar, K. and Sung, C.-J. and Hui, X.},
  journal      = {Fuel},
  volume       = {90},
  number       = {3},
  pages        = {1004--1011},
  year         = {2011},
  doi          = {10.1016/j.fuel.2010.11.022}
}

@article{TARSLiGutmark2005,
  title        = {Effect of exhaust nozzle geometry on combustor flow field and combustion characteristics},
  author       = {Li, G. and Gutmark, E.},
  journal      = {Proceedings of the Combustion Institute},
  volume       = {30},
  pages        = {2893--2901},
  year         = {2005},
  doi          = {10.1016/j.proci.2004.08.189}
}

@article{vauquelin2025TARS,
  title        = {Large eddy simulation of a model jet engine swirl-stabilized flame using sustainable aviation fuels},
  author       = {Vauquelin, P. and Cakir, B. O. and Sanned, D. and Prakash, M. and Hannappel, J.-P. and Subash, A. A. and Richter, M. and Bai, X.-S. and Fureby, C.},
  journal      = {AIAA Paper 2025--0163},
  year         = {January 2025},
  doi          = {10.2514/6.2025-0163}
}

@online{MYTHOS2025,
  title   = {{MYTHOS} Horizon EU — Medium-range hybrid low-pollution flexi-fuel/hydrogen sustainable engine},
  url     = {https://mythos.ruhr-uni-bochum.de/},
  urldate = {2025-12-03}
}

@article{McGuirk_2014, 
    title={The aerodynamic challenges of aeroengine gas-turbine combustion systems}, 
    volume={118},
    DOI={10.1017/S0001924000009386},
    number={1204},
    journal={The Aeronautical Journal}, 
    author={McGuirk, J. J.}, 
    year={2014}, pages={557–599}
}

@misc{Cantera2025,
  author       = {Goodwin, David G. and Moffat, Harry K. and Schoegl, Ingmar and Speth, Raymond L. and Weber, Bryan W.},
  title        = {Cantera: An object-oriented software toolkit for chemical kinetics, thermodynamics, and transport processes},
  year         = {2025},
  howpublished = {\url{https://www.cantera.org}},
  note         = {Version 3.2.0},
  doi          = {10.5281/zenodo.17620923}
}

@misc{EuroHPC,
  title        = {EuroHPC Regular Access Project EHPC-REG-2024R02-188},
  year         = {2024},
  note         = {Supported by the European Union and participating states},
  url          = {https://eurohpc-ju.europa.eu/},
}

\end{document}